\documentclass[preprint,sort&compress,authoryear]{revtex4}
\usepackage{graphicx}
\usepackage{verbatim}
\usepackage{caption}
\usepackage{amsmath}
\usepackage{color}
\usepackage{dirtytalk}
\usepackage[dvipsnames]{xcolor}

\begin{document}

\title{Higher Order Finite Volume Central Schemes for Multi-dimensional Hyperbolic Problems}

\author{Prabal Singh Verma\footnote{prabal.singh.verma@ipp.mpg.de}, 
Wolf-Christian M\"uller}
\affiliation{Technische Universit\"at Berlin, ER 3-2, Hardenbergstr. 36a, 10623 
Berlin, Germany, and Max-Planck/Princeton Center for Plasma Physics}

\begin{abstract}
Different ways of implementing dimension-by-dimension CWENO reconstruction are discussed 
and the most efficient method is applied to develop a fourth order accurate finite volume central scheme 
for multi-dimensional hyperbolic problems.  
Fourth order accuracy and shock capturing nature of the scheme are demonstrated in various nonlinear multi-dimensional problems.
In order to show the overall performance of the present central scheme numerical errors and non-oscillatory behavior 
are compared with existing multi-dimensional CWENO based central schemes for various multi-dimensional problems.
Moreover, the benefits of the present fourth order central scheme over third order implementation are shown by comparing the 
numerical dissipation and computational cost between the two.     

\end{abstract}

\maketitle
\textsl{\textsl{}}
\section{Introduction}
{
Many physical phenomena that exhibit discontinuous solutions can be described by hyperbolic conservation laws. 
Various numerical methods have been 
proposed to approximate the solutions of these hyperbolic conservation laws. Among those, upwind and central schemes 
\cite{lxf,nt,kurganov-2000,kt-2000,kp-2001,knp-2001,lpr1,lpr2,lpr3,lpr4,lpr5,lpr6,lpr7,cs1,cs2,cs3,cs6,cs8,cs10,cs12,cs14,cs15,cs17} 
are widely used examples. An advantage of upwind schemes over central schemes is
{reduced numerical dissipation when dealing with discontinuous solutions. 
Upwind schemes, however, involve the solution of local Riemann problems
\cite{godunov,riemann1,riemann4,riemann5,riemann6} 
which renders such algorithms numerically more complex and more costly. 
On the other hand, central schemes are technically simpler as they do not require Riemann solvers at the price of 
larger numerical diffusion. 
In this paper, we will focus on higher order central schemes in an effort to combine reduced numerical 
dissipation and computational cost.

The very first central scheme was proposed by Lax and Friedrichs \cite{lxf}. 
This scheme employs spatial averaging of neighboring grid cells as part of the integration step --- 
a procedure that can be regarded as an imprecise and highly diffusive partial approximation
of a Riemann problem.
Being first order accurate, the Lax-Friedrichs scheme is numerically too dissipative to be of practical use. 
With the aim of reducing the numerical viscosity, a second order central scheme  
based on a non-oscillatory reconstruction of the linear 
interpolant has been developed \cite{nt}. 
{This approach was further improved} 
by Kurganov et al.\cite{kurganov-2000,kt-2000,kp-2001,knp-2001,kl-2002,kt-2002} by introducing second and third order semi-discrete central schemes. The heart of these
semi-discrete central schemes is centrally weighted essentially non-oscillatory (CWENO) reconstruction of the {local} 
polynomial under consideration. Various third and fourth order CWENO reconstruction methods have been 
proposed for 1D, 2D and 3D hyperbolic conservation laws \cite{lpr1,lpr2,lpr3,lpr4,lpr5,lpr6,lpr7,cs12,cs15}. The CWENO 
method has also been developed for nonuniform meshes and for adaptive mesh refinement \cite{cweno-2016,cweno-weno-2016}. 

Genuine two-dimensional \cite{lpr4}  and three-dimensional \cite{cs12} fourth-order accurate CWENO reconstructions are based on bi-quadratic and triple-quadratic polynomials, respectively.
Hence a truly multidimensional reconstruction requires all cells in a multi-dimensional stencil simultaneously to build up a reconstruction 
polynomial. In contrast, a dimension-by-dimension reconstruction splits  the work into consecutive one-dimensional reconstruction sweeps \cite{weno3}. 
Therefore a genuine multi-dimensional reconstruction is, particularly in three dimensions, 
computationally more expensive than 
a dimension-by-dimension reconstruction. 
Kurganov and Levy \cite{kurganov-2000} have developed a third order semi-discrete central scheme 
for 2D hyperbolic systems using a third order dimension-by-dimension CWENO reconstruction.

Although the dimension-by-dimension approach has been used widely in the WENO framework up to seventh order \cite{weno1,weno2,weno3,weno4,weno5},  
the fourth order CWENO reconstruction \cite{lpr7} has not been investigated using the dimension-by-dimension 
approach for multidimensional problems and hence no comparison (mainly in terms of accuracy and non-oscillatory nature) with truly multidimensional CWENO reconstruction could be made. Moreover, the efficiencies of a third and fourth order CWENO 
reconstructions for multidimensional problems have not been compared yet. 
Hence, the goal of this paper  
is to reconsider these aspects with the aim to provide {for the first time}  
an efficient fourth order finite volume CWENO scheme using a dimension-by-dimension approach.  
It will become clear in the following that the centrally weighted reconstruction is 
particularly beneficial for the construction of an efficient fourth-order central scheme.
 
The flow of this paper is as follows: in section II, we give a brief overview of multidimensional hyperbolic conservation laws and the semi-discrete scheme. 
Section III has been devoted to the fourth order CWENO reconstruction and the dimension-by-dimension 
reconstruction methods for multi-dimensional hyperbolic problems.
Section IV demonstrates the fourth order accuracy of the present scheme by solving various multi-dimensional 
linear and nonlinear hyperbolic problems and provides a comparison with truly multidimensional central schemes. 
In section V non-oscillatory behavior is confirmed and compared with truly multidimensional central schemes 
for the oblique Sod's shock tube problem, oblique Lax problem, 2D blast wave problem, 2D Riemann problem 
3D Burgers' equation and 3D blast wave problem. {
Moreover, Kelvin-Helmholtz instability and shock-bubble interaction  
problems are also solved to demonstrate the 
performance of the scheme for well known more realistic problems.}  
In section VI, we compare the efficiencies of the third-order central scheme \cite{kurganov-2000} with the present fourth order 
central scheme for smooth and non-smooth nonlinear problems. 
In section VII, we provide a brief summary of the work presented in this paper.

\section{Three-dimensional hyperbolic conservation laws and the semi-discrete central scheme}

Three-dimensional hyperbolic conservation laws in general from can be expressed as follows{:}  

 \begin{equation}\label{con}
 \frac{\partial \bf U}{\partial t}+\frac{\partial \bf F^x}{\partial x}+\frac{\partial \bf F^y}{\partial y}+\frac{\partial \bf F^z}{\partial z}=0,%
 \end{equation}

where ${\bf U} \equiv {\bf U}(x,y,z,t)$ is the vector of conserved quantities and ${\bf F^x} \equiv {\bf F^x}\big({\bf U}(x,y,z,t)\big)$, ${\bf F^y} \equiv {\bf F^y}\big({\bf U}(x,y,z,t)\big)$ , 
${\bf F^z} \equiv {\bf F^z}\big({\bf U}(x,y,z,t)\big)$ are the corresponding vectors of fluxes along the $x$-, $y$- and $z$-directions, respectively.  

In order to solve Eq.\eqref{con} numerically we discretize the computational domain $[l_x, l_y, l_z]$ into small grid cells where 
$l_x$, $l_y$, $l_z$ are the lengths of the domain along the $x$-, $y$- and $z$-directions, respectively. Suppose $n_x$, $n_y$ and $n_z$ 
are the number of grid cells along the respective directions then corresponding cell sizes can be 
obtained as follows: 
$\Delta x = l_x/n_x$, $\Delta y = l_y/n_y$ and $\Delta z = l_z/n_z$.  
Now consider a grid cell $(i,j,k)$ centered at ($x_i,y_j,z_k$) and perform the volume integration of Eq.\eqref{con} about the grid cell as, 
 
 \begin{equation}\label{con1}
 \frac{1}{\Delta x\Delta y\Delta z} \\ 
                  \int_{z_{k-1/2}}^{z_{k+1/2}} \int_{y_{j-1/2}}^{y_{j+1/2}} \int_{x_{i-1/2}}^{x_{i+1/2}}\mathrm{d}x  \, \mathrm{d}y \, \mathrm{d}z \,  \Big[ \frac{\partial \bf U}{\partial t}+\frac{\partial \bf F^x}{\partial x}+\frac{\partial \bf F^y}{\partial y}+\frac{\partial \bf F^z}{\partial z}=0 \Big],%
 \end{equation}

here $x_{i\pm1/2} = x_i \pm \Delta x/2$, $y_{j\pm1/2} = y_j \pm \Delta y/2$ and $z_{k\pm1/2} = z_k \pm \Delta z/2$ correspond to the positions of  
grid cell interfaces along the $x$-, $y$- and $z$-directions, respectively. Eq.\eqref{con1} after some algebra becomes,

 \begin{equation}\label{semi-dis}
 \frac{d\overline {\bf U}_{i,j,k}}{d t} = -\frac{\overline{\bf F^x}_{i+1/2,j,k}-\overline{\bf F^x}_{i-1/2,j,k}}{\Delta x}- \\
                              \frac{\overline{\bf F^y}_{i,j+1/2,k}-\overline{\bf F^y}_{i,j-1/2,k}}{\Delta y} -  \\ 
                              \frac{\overline{\bf F^z}_{i,j,k+1/2}-\overline{\bf F^z}_{i,j,k-1/2}}{\Delta z} 
 \end{equation}
 
Here $\overline {\bf U}_{i,j,k}$ is the volume average of ${\bf U}$ in the grid cell $(i,j,k)$ and is defined as,

 \begin{equation}\label{u_avg}
 \overline {\bf U}_{i,j,k}(t) = \frac{1}{\Delta x\Delta y\Delta z} \int_{z_{k-1/2}}^{z_{k+1/2}} \int_{y_{j-1/2}}^{y_{j+1/2}} \int_{x_{i-1/2}}^{x_{i+1/2}}\mathrm{d}x  \, \mathrm{d}y \, \mathrm{d}z \, {\bf U}(x,y,z,t) \, , 
 \end{equation}

and $\overline {\bf F^x}_{i\pm 1/2,j,k}$, $\overline {\bf F^y}_{i,j\pm 1/2,k}$ and $\overline {\bf F^z}_{i,j,k\pm 1/2}$ are the area-averaged fluxes of ${\bf U}$ at the 
grid cell interfaces along the $x$-, $y$- and $z$-directions, respectively and are defined as,

 \begin{equation}\label{F_avg}
 \overline {\bf F^x}_{i\pm 1/2,j,k}(t) = \frac{1}{\Delta y\Delta z} \int_{z_{k-1/2}}^{z_{k+1/2}} \int_{y_{j-1/2}}^{y_{j+1/2}} \mathrm{d}y \, \mathrm{d}z \, 
  {\bf F^x}({\bf U}(x_{i\pm 1/2},y,z,t)) \, , 
 \end{equation}

 \begin{equation}\label{G_avg}
 \overline {\bf F^y}_{i,j\pm 1/2,k}(t) = \frac{1}{\Delta z\Delta x} \int_{x_{i-1/2}}^{x_{i+1/2}}\int_{z_{k-1/2}}^{z_{k+1/2}} \mathrm{d}z \, \mathrm{d}x \, 
  {\bf F^y}({\bf U}(x,y_{j\pm 1/2},z,t)) \, , 
 \end{equation}

 \begin{equation}\label{H_avg}
 \overline {\bf F^z}_{i,j,k\pm 1/2}(t) = \frac{1}{\Delta x\Delta y} \int_{y_{j-1/2}}^{y_{j+1/2}} \int_{x_{i-1/2}}^{x_{i+1/2}}\mathrm{d}x \, \mathrm{d}y \, 
 {\bf F^z}({\bf U}(x,y,z_{k\pm 1/2},t)) \, . 
 \end{equation}

Eq.\eqref{semi-dis} is the finite volume discretization of Eq.\eqref{con} in the semi-discrete form which is evolved in time using 
a classical fourth order Runge-Kutta method \cite{rk4} in order to achieve fourth order accuracy during the temporal evolution. 
Thus, the (spatial) accuracy of the solution is completely determined by the accuracy of the {averaged} fluxes (Eqs.\eqref{F_avg}-\eqref{H_avg}) 
at the grid cell interfaces. However, the fluxes (${\bf F^x},{\bf F^y},{\bf F^z}$) generally (in any nonlinear problem) are  
nonlinear functions of the physical quantities. Therefore, the accuracy of averaged fluxes ($\overline {\bf F^x},\overline {\bf F^y},\overline {\bf F^z}$) depends on the 
accuracy of the point value fluxes which themselves depend on the accuracy of the point value(s) of the physical quantities.

In this work we adopt a dimension-by-dimension CWENO (centrally weighted essentially non-oscillatory) reconstruction to obtain 
fourth order accurate point values of the physical quantities. These CWENO polynomials are reconstructed from the cell averages 
($\overline{\bf U}_{i,j,k}$) so as to recover the point values of the corresponding physical quantities at the grid cell interfaces which 
later are used to compute point value fluxes.      

\section{Dimension by Dimension CWENO Reconstruction}

A dimension-by-dimension third order CWENO approach 
has previously been suggested by Kurganov and Levy \cite{kurganov-2000} for multi-dimensional hyperbolic problems. 
To achieve an accuracy higher than second order one must use 
higher order accurate averaged fluxes because point value fluxes 
would only be second-order approximations and, thus, degrade the accuracy to the 
second order \cite{ppm,weno7}. 
%
      
 {Furthermore}, we emphasize here that 1D fourth order 
CWENO reconstruction \cite{lpr7} has not yet been explored for multi-dimensional problems using the dimension-by-dimension approach.
Moreover, the specific discrete structure of the CWENO technique allows for an efficient implementation of a fourth-order accurate finite-volume scheme.
Therefore, the main objective of the current work is to develop an efficient fourth order accurate finite volume 
CWENO scheme to solve multi-dimensional hyperbolic problems and to provide a comparison with the third order 
CWENO scheme of Kurganov and Levy \cite{kurganov-2000} and with truly multi-dimensional CWENO reconstructions \cite{lpr4,cs12}.

\subsection{Fourth order CWENO reconstruction along the $x$-direction}

The fourth order CWENO reconstruction for 1D hyperbolic problems is comprehensively described by 
Levy et al. \cite{lpr7}.
For the sake of completeness, we give a brief overview of the method.


In each cell $I_{i,j,k}$, one has to reconstruct a quadratic polynomial ${\bf R}_{i,j,k}(x)$ which is a convex 
combination of three quadratic polynomials ${\bf P}_{i-1,j,k}(x)$, ${\bf P}_{i,j,k}(x)$ and ${\bf P}_{i+1,j,k}(x)$ such that, 

\begin{equation} \label{R_I_4th}
{\bf R}_{i,j,k}(x) = \sum_{l = i-1}^{i+1} w_{l,j,k} {\bf P}_{l,j,k}(x),\hspace{0.2cm} \text{where} \hspace{0.2cm} \sum_{l = i-1}^{i+1} w_{l,j,k} = 1
 ,\hspace{0.2cm} w_{l,j,k} \geq 0, \hspace{0.2cm}  \forall \hspace{0.1cm}l \in (i-1, i, i+1).
\end{equation}
The polynomial ${\bf R}_{i,j,k}(x)$ is reconstructed so as to satisfy the three constraints accuracy, cell-average conservation and 
non-oscillatory behaviour. 
The coefficients of the polynomial ${\bf P}_{l,j,k}(x)$ are obtained uniquely by requiring it to conserve the cell averages 
$\overline {\bf U}_{l-1,j,k}$, $\overline {\bf U}_{l,j,k}$ and $\overline {\bf U}_{l+1,j,k}$, where $l \in (i-1,i,i+1)$. Thus, each polynomial, 
${\bf P}_{l,j,k}(x)$, can be written as, 
\begin{eqnarray} \label{P_l}
\nonumber {\bf P}_{l,j,k}(x) = \overline {\bf U}_{l,j,k} - \frac{1}{24}(\overline {\bf U}_{l+1,j,k}-2\overline {\bf U}_{l,j,k}+\overline {\bf U}_{l-1,j,k}) 
+  \frac{\overline {\bf U}_{l+1,j,k} -\overline {\bf U}_{l-1,j,k}}{2 \Delta x}(x-x_l) \\ 
+ \frac{(\overline {\bf U}_{l+1,j,k}-2\overline {\bf U}_{l,j,k}+\overline {\bf U}_{l-1,j,k})} {2\Delta x^2} (x-x_l)^2,\hspace{0.2cm} l = i-1,i,i+1.
\end{eqnarray} 
The nonlinear weights $w_{l,j,k}$ are obtained as, {\it i.e.}
\begin{equation} \label{w_l}
w_{l,j,k} = \frac{\alpha_{l,j,k}}{\alpha_{i-1,j,k}+\alpha_{i,j,k}+\alpha_{i+1,j,k}}, \hspace{0.2cm} 
\text{where} \hspace{0.2cm}\alpha_{l,j,k} = \frac{c_{l,j,k}}{(\epsilon + IS_{l,j,k})^p}, \hspace{0.2cm} \forall \hspace{0.1cm}l \in (i-1,i,i+1).
\end{equation}
Here $\epsilon$, $p$ are chosen to be $10^{-6}$ and $2$, respectively and {the constants $c_{i-1,j,k} = c_{i+1,j,k} = 1/6$, $c_{i,j,k} = 2/3$ 
are chosen so as to guarantee the fourth order accuracy of the point values at the cell-center and cell-boundaries \cite{lpr7}}. $IS_{l,j,k}$ are the 
smoothness indicators which are defined as,
 \begin{equation}\label{ISn}
  IS_{l,j,k} = \sum_{n=1}^2 \int_{x_{i-1/2}}^{x_{i+1/2}} \! (\Delta x)^{2n-1}  ({\bf P}_{l,j,k}^{(n)}(x))^2 \,  \mathrm{d}x , 
\hspace{0.2cm} \forall \hspace{0.1cm}l \in (i-1,i,i+1).
 \end{equation}
Once we have reconstructed all the the polynomials (${\bf P}_{i-1,j,k}$, ${\bf P}_{i,j,k}$, ${\bf P}_{i+1,j,k}$), smoothness indicators can easily be computed using Eq.\eqref{ISn} 
and hence nonlinear weights  
using Eq.\eqref{w_l}. These weights are finally used to reconstruct the polynomial ${\bf R}_{i,j,k}(x)$. 
Please note that we have used the constraint of cell-average conservation to obtain the polynomial ${\bf R}_{i,j,k}(x)$ which is,
 \begin{equation}\label{R_ijk}
 \overline {\bf U}_{i,j,k} = \frac{1}{\Delta x} \int_{x_{i-1/2}}^{x_{i+1/2}}\mathrm{d}x  \,  {\bf R}_{i,j,k}(x)\ . 
 \end{equation}
Now comparing Eq.\eqref{u_avg} and Eq.\eqref{R_ijk} we obtain, 
 \begin{equation}\label{R2_ijk}
{\bf R}_{i,j,k}(x_{i\pm1/2})  = \frac{1}{\Delta y\Delta z} \int_{z_{k-1/2}}^{z_{k+1/2}} \int_{y_{j-1/2}}^{y_{j+1/2}} \mathrm{d}y \, \mathrm{d}z \, {{\bf U}}(x_{i\pm1/2},y,z) \,.   
 \end{equation}
From Eq.\eqref{R2_ijk} it becomes clear that the reconstruction polynomial ${\bf R}_{i,j,k}$ does not 
return point values but area-averaged values at the grid cell interfaces. If we Taylor expand the 
polynomial ${{\bf U}}(x_{i\pm1/2},y,z)$ about the face centers $(x_{i\pm1/2},y_j,z_k)$, 
it can be easily shown that, 
 \begin{eqnarray}\label{R3_ijk}
\nonumber {\bf R}_{i,j,k}(x_{i\pm1/2})  = {{\bf U}}(x_{i\pm1/2},y_j,z_k) 
                                  + \frac{\Delta y^2}{24} 
\partial_{yy}{{\bf U}}(x_{i\pm1/2},y_j,z_k)\\ + \frac{\Delta z^2}{24} 
\partial_{zz}{{\bf U}}(x_{i\pm1/2},y_j,z_k) + {\bf\mathcal O}(\Delta x^4 + \Delta y^4 + \Delta z^4).   
 \end{eqnarray}
This implies that area averages ${\bf R}_{i,j,k}(x_{i\pm1/2})$ are second order  
approximations to the point values at the face centers, {\it i.e.}
 \begin{eqnarray}\label{R4_ijk}
 {\bf R}_{i,j,k}(x_{i\pm1/2})  = {{\bf U}}(x_{i\pm1/2},y_j,z_k) 
                                   + {\bf\mathcal O}(\Delta x^4 + \Delta y^2 + \Delta z^2).   
 \end{eqnarray}
In the next subsection, we discuss different dimension-by-dimension approaches 
to achieve higher order point values and hence higher order averaged fluxes. 
\subsection{Fourth Order Accurate Averaged Fluxes} 
Our ultimate aim is to obtain fourth order accurate averaged fluxes at the grid cell interfaces. Depending on how we compute the averaged flux determines the number of point values which need to be 
reconstructed from the area 
averages ${\bf R}_{i,j,k}(x_{i\pm1/2})$ at the grid cell interfaces along the $x$-direction.

{\bf Method A --} In this case, we apply Simpson's $1/3$ integration over the grid cell interfaces 
and the expression for the averaged flux can thus be written as follows, 
 \begin{eqnarray}\label{Fx_simpson_2d}
\nonumber \overline{\bf F^x}_{i\pm1/2,j,k} =  \frac{1}{36} \Big\{ {\bf f^x}_{i\pm1/2,j-1/2,k-1/2} + 4 {\bf f^x}_{i\pm1/2,j-1/2,k}  
\nonumber                                + {\bf f^x}_{i\pm1/2,j-1/2,k+1/2}  \\+4 ({\bf f^x}_{i\pm1/2,j,k-1/2}  
\nonumber                                + 4 {\bf f^x}_{i\pm1/2,j,k} + {\bf f^x}_{i\pm1/2,j+1/2,k}) 
                                                   \\+{\bf f^x}_{i\pm1/2,j+1/2,k-1/2} + 4 {\bf f^x}_{i\pm1/2,j+1/2,k}   
                                                    + {\bf f^x}_{i\pm1/2,j+1/2,k+1/2} \Big\} + {\bf\mathcal O}(\Delta y^4 + \Delta z^4)
\end{eqnarray} 
Here, {${\bf f^x}$}, is the vector of the fourth order accurate point value fluxes at the 
grid cell interfaces along the $x$-direction and is considered to be a 
simple approximation to the Riemann problem {\it i.e.} the local Lax-Friedrichs flux (LLF). 
The LLF approximation to the point-value flux at the center of  a cell interface is given by,   
 \begin{equation}\label{Fx}
 {\bf f^x}_{i+1/2,j,k} =  \frac{{\bf F^x}({\bf U}_{i+1/2,j,k}^-)+{\bf F^x}({\bf U}_{i+1/2,j,k}^+)}{2} - \\ 
                              \frac{a^x_{i+1/2,j,k}}{2}  ({\bf U}^{-}_{i+1/2,j,k}-{\bf U}^{+}_{i+1/2,j,k})\,, 
 \end{equation}
where the quantities ${\bf U}_{i+1/2,j,k}^+$ and ${\bf U}_{i+1/2,j,k}^-$  are fourth order accurate 
point values at the cell interface as reconstructed from either side of it. They need to be computed from the area averages ${\bf R}_{i,j,k}(x_{i+1/2})$ 
($say$, $\overline{\bf U}_{i,j,k}^+$) and ${\bf R}_{i+1,j,k}(x_{i-1/2})$ ($say$, $\overline{\bf U}_{i+1,j,k}^-$), 
 respectively and ${\bf F^x}({\bf U}_{i+1/2,j,k}^+)$, ${\bf F^x}({\bf U}_{i+1/2,j,k}^-)$ are the respective 
point value fluxes. 
The quantity  $a^x$ 
is the local maximum speed of propagation  
which is estimated as (see for example, \cite{kurganov-2000}),
 \begin{equation}\label{ax}
 a^{x}_{i+1/2,j,k} = \text{max}\left\{ \rho \left(\frac{\partial {\bf F^x}({\bf U}_{i+1/2,j,k}^+)}{\partial U}\right), \\ 
                             \rho \left(\frac{\partial {\bf F^x}({\bf U}_{i+1/2,j,k}^-)}{\partial U}\right) \right\}\,,
 \end{equation}
where $\rho$(A) is the maximum of the magnitude of the eigenvalues of the Jacobian matrix A.
Although, this method has already been used by various authors in solving 2D problems 
(see for example \cite{knp-2001,kl-2002,bal-tad,cs-2005}) using truly two-dimensional 
fourth order CWENO reconstruction, we are applying it dimension-by-dimension in solving 
2D as well as 3D problems.   
Using the 1D CWENO reconstruction polynomial in the subsection {\bf A} we first obtain 
area averages, ${\bf R}_{i,j,k}(x_{i\pm1/2})=\overline{\bf U}_{i,j,k}^{\pm}$ 
at all the grid cell interfaces along the $x$-direction. 
Then we find new 1D CWENO polynomials ${\bf R}_{i,j,k}(y)$ 
based on those area averages 
by following the rules summarized in subsection {\bf A} and adapted to the change from the $x$- to the $y$-direction.
This allows to compute for each $x$-direction cell-interface three equidistant edge-averaged values along 
the $y$-direction, ${\bf R}_{i,j,k}(y_j) = \overline{\bf U}_{i,j,k}^{+0}$ 
($\overline{\bf U}_{i,j,k}^{-0}$), ${\bf R}_{i,j,k}(y_{j\pm1/2}) 
= \overline{\bf U}_{i,j,k}^{+{\pm}}$ ($\overline{\bf U}_{i,j,k}^{-{\pm}}$) - 
one in the face center and other two on the face boundaries, respectively. 
Afterwards, we find new 1D CWENO polynomials ${\bf R}_{i,j,k}(z)$ based on already computed edge-averaged values 
following subsection {\bf A} and adapting to change from the $x$- to the $z$-direction.
This allows, for each $x$-direction cell-interface, to compute three equidistant fourth-order accurate point values along the $z$-direction, one at the edge-center 
${\bf R}_{i,j,k}(z_{k})$ and 
other two at the edge boundaries ${\bf R}_{i,j,k}(z_{k\pm1/2})$. For example, point values 
corresponding to $z$-averaged values $\overline{\bf U}_{i,j,k}^{+0}$ are ${\bf R}_{i,j,k}(z_{k}) = 
{\bf U}_{i+1/2,j,k}^+$ and  ${\bf R}_{i,j,k}(z_{k\pm1/2})  = 
 {\bf U}_{i+1/2,j,k\pm1/2}^+$. Similarly, remaining point values at the grid cell interfaces along the $x$-direction can be 
obtained as shown in figure~\ref{fig1}. 

Thus, in each grid cell we obtain nine fourth order accurate point values (four at the edge corners, 
four at the edge centers and one at the face center as can be seen in figure~\ref{fig1}) on both the grid cell interfaces 
along the $x$-direction. This allows us to obtain the corresponding fourth order accurate 
point value fluxes, ${\bf F^x(U^{\pm})}$ and 
hence the fourth order accurate point value LLF, ${\bf f^x}$, as computed in 
Eqs.\eqref{Fx}-\eqref{ax} for example at ($x_{i+1/2},y_j,z_k$). 
This enables the computation of fourth order accurate 
averaged fluxes $\overline{\bf F^x}$ at all the interfaces along the $x$-direction using Eq.\ref{Fx_simpson_2d}.    

Computation of fourth order accurate averaged fluxes along the $y$- and $z$-direction is 
straightforward due to the simplicity of the dimension-by-dimension approach. For example, 
in order to compute higher order averaged fluxes along the $y$-direction, we first reconstruct a 
new CWENO polynomial ${\bf R}_{i,j,k}(y)$ by adapting to 
change in the direction from $x$ to $y$ in the subsection {\bf A} and 
obtain 
area-averages, ${\bf R}_{i,j,k}(y_{j\pm1/2})=\overline{\bf U}_{i,j,k}^{\pm}$ 
at all the grid cell interfaces along the $y$-direction. Later, these averages are used to reconstruct new 1D 
CWENO polynomials, ${\bf R}_{i,j,k}(z)$ again by adapting to
change in the direction from $x$ to $z$ in the subsection {\bf A} which allow the computation of three equidistant 
edge-averaged values along the $z$-direction, ${\bf R}_{i,j,k}(z_k) = \overline{\bf U}_{i,j,k}^{+0}$
($\overline{\bf U}_{i,j,k}^{-0}$), ${\bf R}_{i,j,k}(z_{k\pm1/2}) 
= \overline{\bf U}_{i,j,k}^{+{\pm}}$ ($\overline{\bf U}_{i,j,k}^{-{\pm}}$) - 
one in the face center and other two on the face boundaries, respectively. Based on these edge-averages
new 1D CWENO polynomial ${\bf R}_{i,j,k}(x)$ is reconstructed so as to obtain corresponding point-values at the 
edge-centers ${\bf R}_{i,j,k}(x_{i})$ and edge boundaries ${\bf R}_{i,j,k}(x_{i\pm1/2})$.   
Thus, we obtain nine fourth order accurate point values 
on both the grid cell interfaces 
along the $y$-direction in each grid cell and this allows us to obtain corresponding fourth 
order accurate point value fluxes, ${\bf F^y(U^{\pm})}$. 
Expression for the  LLF, ${\bf f^y}$ at the face center ($x_i,y_{j+1/2},z_k$) can be obtained 
by adapting to change in the direction from $x$ to $y$ in Eq.\eqref{Fx} as follows,
 \begin{equation}\label{Gy}
 {\bf f^y}_{i,j+1/2,k} =  \frac{{\bf F^y}({\bf U}_{i,j+1/2,k}^-)+{\bf F^y}({\bf U}_{i,j+1/2,k}^+)}{2} - \\ 
                              \frac{a^y_{i,j+1/2,k}}{2}  ({\bf U}^{-}_{i,j+1/2,k}-{\bf U}^{+}_{i,j+1/2,k})\,, 
 \end{equation}
where the quantities ${\bf U}_{i,j+1/2,k}^+$ and ${\bf U}_{i,j+1/2,k}^-$  are the fourth order accurate 
point values computed from the area averages ${\bf R}_{i,j,k}(y_{j+1/2})$ 
 and ${\bf R}_{i,j+1,k}(y_{j-1/2})$, respectively and ${\bf F^y}({\bf U}_{i,j+1/2,k}^+)$, ${\bf F^y}({\bf U}_{i,j+1/2,k}^-)$ 
are the respective point value fluxes. 
And the quantity  $a^y$ 
is the local maximum speeds of propagation  
which is estimated as,
 \begin{equation}\label{ay}
 a^{y}_{i,j+1/2,k} = \text{max}\left\{ \rho \left(\frac{\partial {\bf F^y}({\bf U}_{i,j+1/2,k}^+)}{\partial U}\right), \\ 
                             \rho \left(\frac{\partial {\bf F^y}({\bf U}_{i,j+1/2,k}^-)}{\partial U}\right) \right\}\,,
 \end{equation}
where $\rho$(A) is the maximum of the magnitude of the eigenvalues of the Jacobian matrix A. 
Similarly, remaining point value LLF along the $y$-direction can be obtained.  
Now the expression for the fourth order accurate averaged flux along the $y$-direction can 
also be obtained by adapting to change from $x$- to $y$-direction in Eq.\eqref{Fx_simpson_2d} as, 
 \begin{eqnarray}\label{G_simpson_2d}
\nonumber \overline{\bf F^y}_{i,j\pm1/2,k} =  \frac{1}{36} \Big\{ {\bf f^y}_{i-1/2,j\pm1/2,k-1/2} + 4 {\bf f^y}_{i-1/2,j\pm1/2,k}  
\nonumber                                + {\bf f^y}_{i-1/2,j\pm1/2,k+1/2}  \\+4 ({\bf f^y}_{i,j\pm1/2,k-1/2}  
\nonumber                                + 4 {\bf f^y}_{i,j\pm1/2,k} + {\bf f^y}_{i+1/2,j\pm1/2,k}) 
                                                   \\+{\bf f^y}_{i+1/2,j\pm1/2,k-1/2} + 4 {\bf f^y}_{i+1/2,j\pm1/2,k}   
                                                    + {\bf f^y}_{i+1/2,j\pm1/2,k+1/2} \Big\} + {\bf\mathcal O}(\Delta x^4 + \Delta z^4)
\end{eqnarray} 
Thus, from Eq.\eqref{G_simpson_2d} we obtain the fourth order accurate averaged flux, $\overline{\bf F^y}$, at all 
the interfaces along the $y$-direction. 
Similarly, fourth order accurate averaged flux along the $z$-direction, $\overline{\bf F^z}$  can be easily 
obtained following the above described method }. 
}
{After computing  
fourth order accurate averaged fluxes ($\overline{\bf F^x}$,$\overline{\bf F^y}$,$\overline{\bf F^z}$), Eq.\eqref{semi-dis} 
is evolved using a classical fourth order accurate low-storage Runge-Kutta method \cite{rk4} in order to achieve fourth order accuracy 
during the temporal evolution and the steps are explained as follows: let us assume that 
R.H.S. of Eq.\eqref{semi-dis} is ${\bf C[\overline {U}_{i,j,k}]}$, now dropping the subscripts $(i,j,k)$ Eq.\eqref{semi-dis} can be rewritten as, 
 \begin{equation}\label{semi-dis-new}
 \frac{d\overline {\bf U}(t)}{d t} = {\bf C}[{\bf \overline {U}}(t)] 
 \end{equation}
The intermediate steps to solve Eq.\eqref{semi-dis-new} are as follows, 
 \begin{eqnarray}\label{K1}
  \nonumber    {\bf K}_1 = {\bf C}[{\bf \overline {U}}(t_n)], 
 \end{eqnarray}
 \begin{eqnarray}\label{U1}
   \nonumber   \overline{\bf U}_1 =  {\bf \overline {U}}(t_n) + \frac{\Delta t}{2} {\bf K}_1,
 \end{eqnarray}
 \begin{eqnarray}\label{K2}
   \nonumber   {\bf K}_2 = {\bf C}[{\bf \overline {U}}_1], 
 \end{eqnarray}
 \begin{eqnarray}\label{U2}
  \nonumber    \overline{\bf U}_2 =  {\bf \overline {U}}(t_n) + \frac{\Delta t}{2} {\bf K}_2,
 \end{eqnarray}
  \begin{eqnarray}\label{K3}
   \nonumber   {\bf K}_3 = {\bf C}[{\bf \overline {U}}_2], 
 \end{eqnarray}
 \begin{eqnarray}\label{U3}
   \nonumber   \overline{\bf U}_3 =  {\bf \overline {U}}(t_n) + {\Delta t} {\bf K}_3,
 \end{eqnarray}
  \begin{eqnarray}\label{K4}
   \nonumber   {\bf K}_4 = {\bf C}[{\bf \overline {U}}_3], 
 \end{eqnarray}
 \begin{eqnarray}\label{Un+1}
  \nonumber    \overline{\bf U}(t_{n+1}) =  {\bf \overline {U}}(t_n) + \frac{\Delta t}{6} 
({\bf K}_1+ 2 {\bf K}_2+2 {\bf K}_3+{\bf K}_4).
 \end{eqnarray}
Here, $n = 0, 1, 2, 3,....$ and ${\Delta t}$ is determined dynamically according to the
Courant-Friedrichs-Lewy (CFL) constraint (see for example, \cite{weno-mhd-2016}),  
  \begin{eqnarray}\label{dt}
      \Delta t  = C_{CFL}  min\left(\frac{\Delta x}{a^x_{max}},\frac{\Delta y}{a^y_{max}},\frac{\Delta z}{a^z_{max}}\right), 
 \end{eqnarray}
where, $C_{CFL}$ is the CFL number which for all the 3D tests is chosen as, $0.3$ 
and for all the 2D tests as, $0.45$. The quantities $a^x_{max}$, $a^y_{max}$ 
and $a^z_{max}$ are the maximum values of $a^x_{i+1/2,j,k}$, 
$a^y_{i,j+1/2,k}$ and $a^z_{i,j,k+1/2}$, respectively for all $(i,j,k)$. 

Tests with a modern strongly stability preserving Runge-Kutta scheme of fourth order \cite{SSP10stage} 
have not displayed significant improvements of accuracy but of course {increases} overall performance by allowing larger timesteps.

We have confirmed the fourth order accuracy and the shock capturing 
nature of this method in various nonlinear multidimensional problems. 
However, this method turns out to be computationally too expensive for 3D problems 
as it requires seventy eight sweeps of 1D CWENO reconstruction in each grid cell. 

{\bf Method B --} 
One way of reducing the number of computationally expensive reconstruction steps is to compute the 
averaged flux (say along the $x$-direction) as follows, 
 \begin{eqnarray}\label{Fx_2d}
\nonumber {\bf \overline F^x}_{i\pm1/2,j,k} =   {\bf f^x}_{i\pm1/2,j,k} +\frac{1}{24} ( {\bf f^x}_{i\pm1/2,j-1,k}  
\nonumber                                - 2 {\bf f^x}_{i\pm1/2,j,k} + {\bf f^x}_{i\pm1/2,j+1,k}) \\ 
                                           +\frac{1}{24} ( {\bf f^x}_{i\pm1/2,j,k-1}  
                                - 2 {\bf f^x}_{i\pm1/2,j,k} + {\bf f^x}_{i\pm1/2,j,k+1}) + {\bf\mathcal O}(\Delta y^4 + \Delta z^4) .
 \end{eqnarray}
This approximation to the averaged flux can be obtained from Eq.\eqref{R3_ijk} by replacing ${\bf R}_{i,j,k}(x_{i\pm1/2})$ 
(area-average quantity) 
with ${\bf \overline F^x}_{i\pm1/2,j,k}$, ${{\bf U}}(x_{i\pm1/2},y_j,z_k)$ (point value at the face-center) 
with ${\bf f^x}_{i\pm1/2,j,k}$ and second order derivatives along the $y$- and $z$-directions are approximated by 
finite difference formula for point values at the face-center. Thus, in this method as opposed to Eq.\eqref{Fx_simpson_2d} we just need to 
obtain point 
values at the face centers. This averaging procedure has previously been used to obtain higher order 
averaged flux in the multidimensional PPM (piecewise parabolic method) \cite{ppm} and WENO schemes \cite{weno7}. 
Whereas, we are applying it in the framework of a central scheme which naturally allows the computation of 
non-oscillatory point values (${\bf U}_{i+1/2,j,k}^+$, ${\bf U}_{i-1/2,j,k}^-$) at the face-center. 
The procedure for computing fourth order accurate point values at the face-centers and the corresponding point value LLF (Eq.\eqref{Fx}) 
is already described for method A. Once we know the point value LLF fluxes at the face centers, ${\bf f^x}_{i\pm1/2,j,k}$, 
the averaged fluxes along the $x$-direction, ${\bf \overline F^x}_{i\pm1/2,j,k}$ can straightforwardly be obtained from Eq.\eqref{Fx_2d}. 
The expression for the fourth order accurate averaged flux along the $y(z)$-direction, $\overline{\bf F^y}(\overline{\bf F^z})$, can
be obtained by adapting to the change from the $x$-direction to the $y(z)$-direction in Eq.\eqref{Fx_2d}. 
After the computation of   
the fourth order accurate averaged fluxes ($\overline{\bf F^x}$,$\overline{\bf F^y}$,$\overline{\bf F^z}$), Eq.\eqref{semi-dis} 
is evolved using a fourth order accurate Runge-Kutta method as described for method A. 

We have confirmed the fourth order accuracy and the shock capturing 
nature of this procedure as well in various nonlinear multidimensional problems. 
Moreover, this method turns out to be computationally lesser expensive  
as it requires only twelve sweeps of 1D CWENO reconstruction in each grid cell for 3D problems.   

{\bf Method C --}  one can further reduce the computational cost by obtaining the fourth 
order accurate point value at the face center as follows, 
 \begin{eqnarray}\label{ue_2d}
\nonumber {{\bf U}}_{i+1/2,j,k}^+ =   \overline{\bf U}_{i,j,k}^+ -\frac{1}{24} ( \overline{\bf U}_{i,j-1,k}^+  
\nonumber                                - 2 \overline{\bf U}_{i,j,k}^+ + \overline{\bf U}_{i,j+1,k}^+) \\ 
                                           -\frac{1}{24} ( \overline{\bf U}_{i,j,k-1}^+  
                                - 2 \overline{\bf U}_{i,j,k}^+ + \overline{\bf U}_{i,j,k+1}^+) + {\bf\mathcal O}(\Delta y^4 + \Delta z^4). 
 \end{eqnarray}
This expression can also be derived from Eq.\eqref{R3_ijk} by replacing ${\bf R}_{i,j,k}(x_{i+1/2})$ 
(area-average quantity) 
with ${\bf \overline U}_{i,j,k}^+$, ${{\bf U}}(x_{i+1/2},y_j,z_k)$ 
(point value at the face-center) 
with ${\bf U}_{i+1/2,j,k}^+$ and by approximating second order derivatives along 
the $y$- and $z$-directions using the finite difference formula for area-averages, 
${\bf \overline U}_{i,j,k}^+$. Similarly, one can obtain the point value at the center of the 
opposite interface, ${\bf U}_{i-1/2,j,k}^-$ by replacing $\overline{\bf U}_{i,j,k}^+$ with $\overline{\bf U}_{i,j,k}^-$ 
and ${{\bf U}}_{i+1/2,j,k}^+$ with ${{\bf U}}_{i-1/2,j,k}^-$ in Eq.\eqref{ue_2d} as,
 \begin{eqnarray}\label{uw_2d}
\nonumber {{\bf U}}_{i-1/2,j,k}^- =   \overline{\bf U}_{i,j,k}^- -\frac{1}{24} ( \overline{\bf U}_{i,j-1,k}^-  
\nonumber                                - 2 \overline{\bf U}_{i,j,k}^- + \overline{\bf U}_{i,j+1,k}^-) \\ 
                                           -\frac{1}{24} ( \overline{\bf U}_{i,j,k-1}^-  
                                - 2 \overline{\bf U}_{i,j,k}^- + \overline{\bf U}_{i,j,k+1}^-) + {\bf\mathcal O}(\Delta y^4 + \Delta z^4). 
 \end{eqnarray}
This method of computing higher-order point value at the face-center has previously been used in the framework of 
the third order CWENO scheme of Kurganov and Levy \cite{kurganov-2000}, fourth order PPM scheme \cite{ppm} and 
fourth order WENO scheme \cite{weno7}. We are applying this method to develop a fourth-order accurate CWENO scheme.  
The expressions for the fourth-order accurate point values at the face-centers along the $y(z)$-direction 
can be obtained from Eqs.\eqref{ue_2d}-\eqref{uw_2d} by adapting to the change from $x$-direction to $y(z)$-direction.
Once the point values at all the face-centers are computed, averaged fluxes ($\overline{\bf F^x}$,$\overline{\bf F^y}$,$\overline{\bf F^z}$) 
  are obtained as described for method A.  The temporal evolution of 
Eq.\eqref{semi-dis} is same as described for method A.  
 
We have compared the accuracy, non-oscillatory behavior and computational expense of all the three methods for 
various multi-dimensional nonlinear problems and found no significant difference among the three except the 
computational cost.   

In TABLE \ref{tab1}, we compare the performance of the three methods for 3D linear advection test and find that the 
method C is approximately $4.3$ times 
faster than method A and $1.87$ times faster than method B.

\begin{center}
\begin{tabular}{ |p{1.5cm}|p{3cm}|p{3cm}|p{3.1cm}| }
 \hline
NG & Method C & Method A & Method B \\
 \hline
$32^3$ & 1.0  & 6.06 & 1.87\\
$64^3$ & 1.0 &  3.57 & 1.93\\
$128^3$ &  1.0  & 3.46 & 1.88\\
$256^3$& 1.0  & 3.32 & 1.95\\
$512^3$& 1.0  & 5.09 & 1.72\\
 \hline
Mean & 1.0  & 4.3 & 1.87\\

 \hline
\end{tabular}
\captionof{table}{Performance test for the three methods where the run time is normalized with respect 
to method C} \label{tab1}
\end{center}
Since method C seems to be the most efficient in terms of computational cost, therefore, in this manuscript we will show robustness of  
this method by solving various multidimensional problems. 
Please note here that although with this method we do not encounter the problem of 
negative pressure or density in any of the tests while solving Euler equations, it can not be ruled out that strong discontinuities may introduce 
such unphysical effects. 
Therefore, we recommend to check for the positivity of density and pressure after the 
computation of point values in Eqs.\eqref{ue_2d}-\eqref{uw_2d} and switch off the 
addition of approximated derivatives when the check fails. Thus, in such situations the point value at the 
face center, ${{\bf U}}_{i+1/2,j,k}^+$ is replaced by its second order approximation, ${\overline{\bf U}}_{i,j,k}^+$. 
%
%
%
%
%
%
}

{
It is to note here that 3D hyperbolic equations reduce to 2D hyperbolic equations when there are no variations along the $z$-direction. 
Therefore, a 2D hyperbolic equation is a simpler version of a 3D equation \eqref{con} which can be obtained by 
dropping the variations along the $z$-direction {\it i.e.}, the fourth term $(\frac{\partial \bf F^z}{\partial z})$ in Eq.\eqref{con}. 
Thus, in 2D problems volume-averaged quantities ($say, \overline{\bf U}_{i,j,k}$) 
would reduce to area-averages ($\overline{\bf U}_{i,j}$) 
and area-averaged quantities ($say, \overline {\bf F^x}_{i\pm 1/2,j,k}$) would reduce to edge-averages $(\overline {\bf F^x}_{i\pm 1/2,j})$. 
So, the discretization and the numerical schemes for 2D problems can be recovered just by dropping the variation along the 
$z$-direction as well as the subscripts ``$k,k\pm$" in all the expression throughout the paper.  
}

The code is MPI (message-passing-interface) parallel, using a two-dimensional decomposition 
of the computational grid. The 2D decomposition is also referred to as the pencil 
decomposition or column decomposition. It is used when
the number of processes comes close to, or even exceeds, the number of grid cells in one spatial dimension.
In two of the three space dimensions the grid is divided and distributed over the processes. The third
dimension resides entirely within each process. All tests have been performed on a parallel architecture based on Intel Xeon `Sandybridge' cpus.


\section{Accuracy}

In this section, we present the convergence of errors by solving the Euler equations of gas dynamics. 
 We first extract the 1D profile of a physical quantity from the 2D (or 3D) grid and then  
the norm of the error for all the convergence studies can be computed as follows, 

\begin{equation}
 L_1 = \frac{1}{NG}\sum_{i = 1}^{NG} |E_i^f - E_i^0|,  
\end{equation} 
where $E_i^0$, $E_i^f$ are the exact reference and the numerical solutions as a function of grid resolution and 
$NG$ is the number of grid points. 

After computing the norms of the errors, we obtain the experimental order
of convergence $(EOC)$ using the formula,
\begin{equation}
  EOC(j) = \frac{|log(L_1(NG(j)))|-|log(L_1(NG(j-1)))|}{|log(NG(j))|-|log(NG(j-1))|},   
\end{equation}

here $j$ runs over the indices of the  column vectors in the tables shown in the later subsections.

\subsection{The Euler Equations of Gas Dynamics}
Equations governing the dynamics of a 3D adiabatic system can be described by Eq.\eqref{con} where,
\begin{eqnarray}\label{uavg}
{\bf U} = \left( \begin{array}{c} \rho\\ \rho v_x\\\rho v_y\\\rho v_z\\  E\end{array}\right), \hspace{0.2cm}  
{\bf F^x} = \left( \begin{array}{c} \rho v_x\\ \rho v_x^2+p\\\rho v_x v_y\\\rho v_x v_z\\ v_x(E+p) \end{array}\right), \hspace{0.2cm} 
{\bf F^y} = \left( \begin{array}{c} \rho v_y\\ \rho v_x v_y\\\rho v_y^2+p\\\rho v_y v_z\\ v_y(E+p) \end{array}\right), \hspace{0.2cm} 
{\bf F^z} = \left( \begin{array}{c} \rho v_z\\ \rho v_x v_z\\\rho v_y v_z\\ \rho v_z^2+p\\v_z(E+p) \end{array}\right). 
\end{eqnarray}
Here $\rho$ is the density, $v_x$, $v_y$, $v_z$ are $x$, $y$ and $z$ component of the velocity, $E$ is the total 
energy and $p$ is the pressure which is related to $E$ through $p = (\gamma -1)(E-0.5 \rho (v_x^2+v_y^2+v_z^2))$. 
One may recover 2D adiabatic systems just by dropping the $z$ components from Eq. \eqref{uavg}.  
Boundary conditions are periodic.

\subsubsection{Linear Problems}

$Example$ $1.$ The initial conditions, for the 2D and 3D cases, are chosen as follows:

\vspace{-1cm}

 \begin{eqnarray}\label{sound2d}
\nonumber \rho(x,y,0) = 1 + 0.5 \sin [2\pi(x/l_x + y/l_y)], \\ 
\nonumber v_x(x,y,0)  = 1 = v_y(x,y,0),  \\
 \nonumber p(x,y,0) = 3/5, \hspace{0.2cm} l_x\times l_y = [0,1] \times [0,1]. 
 \end{eqnarray}

\vspace{-1cm}

 \begin{eqnarray}\label{sound3d}
\nonumber  \rho(x,y,z,0) = 1 + 0.5 \sin [2\pi(x/l_x + y/l_y + z/l_z)], \\
 \nonumber v_x(x,y,z,0)  = 1.0 = 
 \nonumber v_y(x,y,z,0)  = v_z(x,y,z,0) , \\ 
 \nonumber p(x,y,z,0) = 3/5,\hspace{0.2cm}  l_x \times l_y \times l_z  =[0,1] \times [0,1] \times [0,1].
 \end{eqnarray}
Here $\gamma$ is chosen to be $5/3$ and the numerical solutions are obtained after one period. 
These initial conditions result in the advection of the initial density profile due to constant velocity components and constant pressure
(see for example \cite{weno7}). 
TABLE \ref{tab2} and TABLE \ref{tab3} contain the convergence of errors for the 2D and 3D 
linear Euler gas dynamics, respectively.
\begin{center}
\begin{tabular}{ |p{1cm}|p{1cm}|p{3cm}|p{1cm}| }
 \hline
j & NG & $L_1$ & EOC   \\
 \hline
1 & $16^2$   & 1.209E-3    & - \\
2 & $32^2$ &   4.367E-5  & 4.79 \\
3 & $64^2$ &   1.616E-6 & 4.75 \\
4 & $128^2$  & 7.413E-8  & 4.44 \\
5 & $256^2$&   4.075E-9  & 4.18 \\
 \hline
\end{tabular}
\captionof{table}{Convergence of errors for the 2D linear Euler gas dynamics} \label{tab2}
\end{center}
\begin{center}
\begin{tabular}{ |p{1cm}|p{1cm}|p{3cm}|p{1cm}| }
 \hline
 j &NG & $L_1$ & EOC   \\
 \hline
1 & $16^3$   & 4.793E-3    & - \\
2 & $32^3$ &   1.753E-4  & 4.77 \\
3 & $64^3$ &   6.389E-6 & 4.77 \\
4 & $128^3$  & 2.876E-7  & 4.47 \\
5 & $256^3$&   1.637E-8  & 4.13 \\
 \hline
\end{tabular}
\captionof{table}{Convergence of errors for the 3D linear Euler gas dynamics} \label{tab3}
\end{center}

\subsubsection{Nonlinear Problems}

$Example$ $2.$ Initial conditions for the 2D Euler vortex evolution problem (see for example \cite{weno1}) are, 
\begin{eqnarray}\label{vortex}
\nonumber \left( \begin{array}{c} \rho\\ v_x\\ v_y\\ p\end{array}\right) = \left( \begin{array}{c} (1 + \delta T)^{1/(\gamma-1)}\\ 
                      1 + (l_y/2 - y) \frac{\sigma}{2 \pi} e^{0.5 (1-r^2)}\\ 1 + (x - l_x/2) \frac{\sigma}{2 \pi} e^{0.5 (1-r^2)}\\(1 + \delta T)^{\gamma/(\gamma-1)}\end{array}\right),
 \end{eqnarray} 
Here $\delta T$ is the perturbation in the temperature and is given by,
\begin{eqnarray}\label{temp}
\nonumber\delta T = - \frac{(\gamma-1)\sigma^2}{8 \gamma \pi^2} e^{(1-r^2)},
\end{eqnarray}
where  $r^2 = (x-l_x/2)^2 +(y-l_y/2)^2  $ and the vortex strength $\sigma$ = 5,  $l_x \times l_y  = [0,10] \times [0,10]$. 
Here $\gamma$ is chosen to be $1.4$. The initial conditions lead advection of a non-linear vortex at an angle of $45^\circ$ with the $x$-axis and 
the numerical solutions are obtained after one period ($t = 10$).
Convergence of errors for this nonlinear problem are presented in TABLE \ref{tab4}.  
\begin{center}
\begin{tabular}{ |p{1cm}|p{1cm}|p{3cm}|p{1cm}| }
 \hline
j & NG & $L_1$ & EOC   \\
 \hline
1 & $32^2$   & 6.584E-3    & - \\
2 & $96^2$ &   5.255E-5  & 4.39 \\
3 & $160^2$ &  4.449E-6  & 4.83 \\
4 & $288^2$  & 2.757E-7  & 4.73 \\
 \hline
\end{tabular}
\captionof{table}{2D vortex evolution problem} \label{tab4}
\end{center}

$Example$ $3.$ Euler equations with the initial conditions (see for example \cite{weno7}),  
 \begin{eqnarray}\label{euler2d}
\nonumber \rho(x,y,0) = 1 + 0.5 \sin [\pi(x + y - 2)], \\ 
\nonumber v_x(x,y,0)  = \cos [\pi(x + 2 y - 3)], 
 \nonumber v_y(x,y,0) = 1 - 0.5 \sin [\pi(2 x + y - 3)], \\
 \nonumber p(x,y,0) = 1 - 0.5 \sin [\pi(x - y)], \hspace{0.2cm} l_x \times l_y = [0,2] \times [0,2]. 
 \end{eqnarray}
follow the nonlinear evolution and lead to discontinuous solution after a finite time.
Here $\gamma$ is chosen to be $1.4$ and the numerical solutions are obtained at $t = 0.05$ 
(solution is still smooth) and the reference solution is obtained at high resolution $(1056 \times 1056)$. 
Convergence of errors for this nonlinear problem are shown in TABLE \ref{tab5}.  
\begin{center}
\begin{tabular}{ |p{1cm}|p{1cm}|p{3cm}|p{1cm}| }
 \hline
j & NG & $L_1$ & EOC   \\
 \hline
1 & $32^2$   & 4.686E-4    & - \\
2 & $96^2$ &   6.068E-6  & 3.96 \\
3 & $160^2$ &  4.459E-7  & 5.11 \\
4 & $288^2$  & 2.821E-8  & 4.69 \\
 \hline
\end{tabular}
\captionof{table}{2D nonlinear Euler problem} \label{tab5}
\end{center}
{  
\subsection{Comparison with truly multidimensional CWENO schemes}
In this section we compare the convergence of errors for the 3D linear advection test and 3D Burgers' equation 
with a truly 3D central scheme \cite{cs12}.  

$Example$ $4.$ Three dimensional linear advection test -- 

Equation for the 3D linear advection test is,
\begin{eqnarray}\label{temp}
 \nonumber \frac{\partial U}{\partial t}+\frac{\partial U}{\partial x}+\frac{\partial U}{\partial y}+\frac{\partial U}{\partial z}=0,%
\end{eqnarray}
and the initial condition is chosen to be,
\begin{eqnarray}\label{temp}
 \nonumber U(x,y,z) = \sin^2(\pi x)\sin^2(\pi y)\sin^2(\pi z).%
\end{eqnarray}
Computational domain is a unit cube, {\it i.e.},  $l_x \times l_y \times l_z= [0,1] \times [0,1] \times [0,1]$ and 
the boundary conditions are periodic. Thus, all the parameters for this test are chosen to be same as in ref. \cite{cs12}. In the TABLE \ref{tab6}  we 
provide a comparison for the error norm,   
\begin{center}
\begin{tabular}{ |p{1cm}|p{1cm}|p{3cm}|p{1cm}|p{3.2cm}|p{1cm}| }
 \hline
j & &
\multicolumn{2}{|c|}{dim-by-dim}& 
\multicolumn{2}{|c|}{3D CWENO} \\ 
\cline{3-6}
 & NG & $L_1$ & EOC  & $L_1$ & EOC   \\
 \hline
1 & $10^3$   & 2.252E-2    & - &  1.08E-2     & -\\ 
2 & $20^3$ &  9.728E-4   & 4.53 & 6.395E-4   & 4.09 \\
3 & $40^3$ &  4.024E-5  & 4.59 &  3.834E-5   & 4.06 \\
4 & $80^3$  & 1.875E-6  & 4.42 &  2.3633E-6   &4.02 \\
 \hline
\end{tabular}
\captionof{table}{3D linear advection test} \label{tab6}
\end{center}

$Example$ $5.$ Three dimensional Burgers' equation -- 

Equation for the 3D Burgers' equation is expressed as follows,
\begin{eqnarray}\label{temp}
 \nonumber \frac{\partial U}{\partial t}+\frac{\partial }{\partial x}\Big(\frac{U^2 }{2}\Big)+\frac{\partial }{\partial y}\Big(\frac{U^2 }{2}\Big)
+\frac{\partial }{\partial z}\Big(\frac{U^2 }{2}\Big)=0,%
\end{eqnarray}
and the initial condition is chosen to be,
\begin{eqnarray}\label{temp}
 \nonumber U(x,y,z) = 0.25 + \sin(\pi x)\sin(\pi y)\sin(\pi z).%
\end{eqnarray}
Computational domain is chosen as, $l_x \times l_y \times l_z= [-1,1] \times [-1,1] \times [-1,1]$ and 
the boundary conditions are periodic. Thus, all the parameters for this test also are chosen to be same as in ref. \cite{cs12}.  
The only difference is that we obtain the reference solution by evolving the equation at high resolution ($510^3$).   
 In the TABLE \ref{tab7}  we 
provide a comparison for $L_1$ error norm,   

\begin{center}
\begin{tabular}{ |p{1cm}|p{1cm}|p{3cm}|p{1cm}|p{1cm}|p{3.2cm}|p{1cm}| }
 \hline
j &
\multicolumn{3}{|c|}{dim-by-dim}& 
\multicolumn{3}{|c|}{3D CWENO} \\ 
\cline{2-7}
 & NG & $L_1$   & EOC  &  NG & $L_1$  & EOC  \\
 \hline
1 & $10^3$   & 4.005E-3    & -  & $10^3$   & 6.846E-3    & -  \\ 
2 & $30^3$ &   5.915E-5  & 3.84 & $20^3$ &   5.804E-4  & 3.56  \\
3 & $50^3$ &   6.692E-6 & 4.27 & $40^3$ &   3.347E-5 & 4.16   \\
4 & $90^3$  &  5.079E-7 & 4.39 & $80^3$  &  1.766E-6 & 4.21   \\
 \hline
\end{tabular}
\captionof{table}{3D Burgers' equation} \label{tab7}
\end{center}


\section{Non-oscillatory behavior}
 
In this section we present solutions for the oblique Sod's shock tube problem, the oblique Lax problem, the 2D blast wave, 
the 2D Riemann problem, the 3D Burgers' equation and the 3D blast wave problem  
to demonstrate the 
non-oscillatory behavior and compare the results with genuine multi-dimensional fourth order accurate central schemes 
\cite{lpr4,cs12}. Moreover, well known and more realistic problems like 
Kelvin-Helmholtz instability \cite{KH} and shock-bubble interaction \cite{shock-cloud2D,shock-cloud3D} problems are also solved to demonstrate the 
robustness of the scheme.     

{\bf The oblique Sod's shock tube problem} -- We solve the Euler gas dynamics, described in Eq.\eqref{uavg}, for a 
1D shock tube initial value problem \cite{sod} on 2D grid such that the initial discontinuity makes an angle of $60^{\circ}$ with the $x$-axis. 
Initial condition for this test are chosen to be same as in ref. \cite{lpr4} and can be expressed as,
\begin{eqnarray}
\nonumber \left( \begin{array}{c} \rho\\ v_x\\ v_y \\p\end{array}\right)_L = \left( \begin{array}{c} 1\\ 0\\ 0\\ 1\end{array}\right), \hspace{0.8cm}  
\left( \begin{array}{c} \rho\\ v_x\\ v_y \\p\end{array}\right)_R = \left( \begin{array}{c} 0.125\\ 0\\ 0\\ 0.1\end{array}\right),   
\end{eqnarray}
In this test $\gamma$ is chosen to be $1.4$.  
The computational domain is the rectangle, $l_x \times l_y = [-0.5,0.5]\times[-0.125,0.125]$ and boundary conditions are open.  
The solution is sampled at $y = 0.0$  at time $t = 0.16632$. In figure~\ref{fig2}, we compare the solutions at two different resolutions 
$[200\times50]$  and $[600\times150]$. At the coarser resolution $[200\times50]$, we see some small amplitude wiggles, 
near the contact discontinuity (also observed in reference \cite{lpr4}) which gets smaller at a finer resolution $[600\times150]$. 

{\bf The oblique Lax problem} --  In this problem we again solve the Euler gas dynamics, described in Eq.\eqref{uavg}, 
on the 2D grid such that the initial discontinuity makes an angle of $45^{\circ}$ with the $x$-axis and the initial conditions are expressed as, 
\begin{eqnarray}
\nonumber \left( \begin{array}{c} \rho\\ v_x\\ v_y \\p\end{array}\right)_L = \left( \begin{array}{c} 0.445\\ 0.698/\sqrt 2\\ 0.698/\sqrt 2\\ 3.528\end{array}\right), \hspace{0.8cm}  
\left( \begin{array}{c} \rho\\ v_x\\ v_y \\p\end{array}\right)_R = \left( \begin{array}{c} 0.5\\ 0\\ 0\\ 0.571\end{array}\right).   
\end{eqnarray}
Here $\gamma$ is chosen to be $1.4$, computational domain as, $l_x \times l_y = [-0.5,0.5]\times[-0.5,0.5]$ and boundary conditions are open.  
 In figure~\ref{lax}, two solutions at the resolutions $200^2$ and $600^2$ are 
compared at time $t = 0.12$ and in this test also it is found that, small amplitude wiggles near the contact discontinuity 
get smaller at a finer resolution and get further smaller with time.  

{Please note here that changing the time integrator for example low storage strong stability-preserving Runge-Kutta method \cite{SSP10stage}
does not affect the amplitude of the wiggles. 
}

{\bf The 2D blast wave problem} -- Here we solve Eq.\eqref{uavg} for the same initial condition as in \cite{lpr4} which can 
be expressed as follows, 
 \begin{eqnarray}
  \nonumber \left( \begin{array}{c} \rho, v_x,  v_y,  p\end{array}\right) = \left \{ \begin{array}{c} (1, 0, 0,  1) \hspace{0.2cm} 
  \text{if}  \hspace{0.2cm} (x-l_x/2)^2 + (y-l_y/2)^2 \le R^2 \\ (0.125, 0, 0, 0.1)   \hspace{0.2cm} 
  \text{otherwise}  \hspace{2.8cm} \end{array}\right \}.  
  \end{eqnarray}
Here $R = 0.2$, $\gamma$ = $1.4$, the boundary conditions are periodic and the computational domain is chosen as, $l_x \times l_y = [0,1]\times[0,1]$. We compare the 
solutions, at time $t = 0.1$ at grid resolutions $100^2$ and $200^2$, 
shown in figure~\ref{fig3} and figure~\ref{fig4}, respectively. 
From these figures it becomes clear that our scheme in spite of being so simple produces
results which are very close to truly multidimensional CWENO scheme \cite{lpr4} as the symmetry loss gets better 
at higher resolution. Probably, using more accurate numerical flux 
(HLL, see for example \cite{riemann6}) may reduce the symmetry loss.  

{\bf The 2D Riemann problem} -- In this test, we solve Eq.\eqref{uavg} for configuration $5$ of ref.\cite{kt-2002} which is also used in 
ref.\cite{lpr4}, so the initial conditions are chosen as, 
 \begin{eqnarray}
  \nonumber \left( \begin{array}{c} \rho, v_x,  v_y,  p\end{array}\right) = \left \{ \begin{array}{c} (1, -0.75, -0.5,  1) \hspace{0.2cm} 
  \text{if}  \hspace{0.2cm} x > 0, y > 0  \\ (2, -0.75, 0.5, 1)  \hspace{0.2cm} 
  \text{if}  \hspace{0.2cm} x < 0, y > 0 
   \\ (1, 0.75, 0.5, 1)   \hspace{0.2cm} 
  \text{if}  \hspace{0.2cm} x < 0, y < 0   \\ (3, 0.75, -0.5, 1)   \hspace{0.2cm} 
  \text{if}  \hspace{0.2cm} x > 0, y < 0  \end{array}\right \}.  
  \end{eqnarray}
 Here $\gamma$ = $1.4$, the boundary conditions are open and the computational domain is chosen as, $l_x \times l_y = [-0.5,0.5]\times[-0.5,0.5]$. 
This configuration results in four interacting contact discontinuities. 
The solutions are obtained at $t = 0.23$ at two different grid resolutions $200^2$  and $400^2$ shown in 
figure~\ref{fig5} and figure~\ref{fig6}, respectively. For the adiabatic system considered here, the obtained solution is not expected to be symmetric about the oirigin \cite{schulz1993numerical}.  
This test displays the scheme's ability to correctly handle the partly intricate structuring in spite of the method's comparable simplicity.  

{\bf The 3D Burgers' equation} -- The problem of the 3D Burgers' equation is already described in $Example$ $5$ 
and solutions are obtained at time $t = 0.8$.  
Figure~\ref{fig7} and figure~\ref{fig8} show the contour plots at resolutions $80^3$ (same as in ref.\cite{cs12}) 
and $400^3$, respectively in the plane $z = 0$.  
}

{\bf The 3D blast (explosion) test problem} --  In this problem we solve 3D Euler equation [Eq.\eqref{uavg}] 
for the same initial conditions as used in ref. \cite{cs12} which are expressed as follows, 
 \begin{eqnarray}
  \nonumber \left( \begin{array}{c} \rho, v_x,  v_y, v_z, p\end{array}\right) = \left \{ \begin{array}{c} (1, 0, 0, 0, 1) \hspace{0.2cm} 
  \text{if}  \hspace{0.2cm} (x-l_x/2)^2 + (y-l_y/2)^2  + (z-l_z/2)^2 \le R^2 \\ (0.125, 0, 0, 0, 0.1)   \hspace{0.2cm} 
  \text{otherwise}  \hspace{5.5cm} \end{array}\right \}.  
  \end{eqnarray}
Here $R = 0.2$, $\gamma$ = $1.4$, the boundary conditions are periodic and the computational domain is unit-cube, $l_x \times l_y \times l_z = [0,1]\times[0,1]\times[0,1]$.
Solutions are obtained at $t = 0.1$. Figure~\ref{fig9} shows the 1D profile of the density $(\rho)$, obtained along the 
intersection of planes $y = 0.5$ and $z = 0.5$ at a resolution $80^3$ (same as in ref.\cite{cs12}) 
and the reference solution is obtained at a high resolution $720^3$.  

{\bf Kelvin-Helmholtz (KH) instability} --  In this problem we solve Eq.\eqref{uavg} in 2D along with periodic 
boundary conditions and the computational domain is set as, $l_x \times l_y = [-0.5,0.5]\times[-0.5,0.5]$. The pressure is 
in equilibrium with $p = 2.5$ everywhere and $\gamma = 5/3$. Here the problem is initialized as suggested in ref. 
\cite{KH}, so the initial conditions can be expressed as,
 \begin{eqnarray}\label{kh}
   \left( \begin{array}{c} \rho, v_x,  v_y,  p\end{array}\right) = \left \{ \begin{array}{c} (2, 0.5, 0, 2.5) \hspace{0.2cm} 
  \text{if}  \hspace{0.2cm} |y|< 0.25  \\ (1, -0.5, -0.025 \sin [2\pi(x+0.5) / \lambda ], 2.5)  \hspace{0.2cm} 
  \text{if}  \hspace{0.2cm} |y-0.25| < 0.025  
   \\ (1, -0.5, 0.025 \sin[2\pi(x+0.5) / \lambda], 2.5)  \hspace{0.2cm} 
  \text{if}  \hspace{0.2cm}  |y+0.25| < 0.025  \\ (1, -0.5, 0, 2.5)   \hspace{0.2cm} 
  \text{elsewhere }  \hspace{0.2cm}  \end{array}\right \}.  
  \end{eqnarray}
Here, wavelength of the perturbation, $\lambda$ is chosen to be $6$. Figure~\ref{fig-kh-1024} shows the snap-shots of color-coded contour plots for the density at time $t = 0.7, 1.4, 2.1, 2.8$ 
at a resolution $1024^2$. {Generation of small scale structures during the nonlinear development of the KH instability are symmetrically captured.} 

{\bf Shock-Bubble Interaction} -- In this test, Eq.\eqref{uavg} is solved to show the interaction 
of a planar shock with a low density circular region for the 2D test \cite{shock-cloud2D} 
and with low density spherical region for the 3D test \cite{shock-cloud3D}. The computational domain for the 
2D and 3D cases are chosen as, $l_x \times l_y = [-0.1,1.5]\times[-0.5,0.5]$ with resolution $400^2$ and $l_x \times l_y \times l_z = [-0.1,1.5]\times[-0.5,0.5]\times[-0.5,0.5]$ 
with resolution $400^3$, respectively. Boundary conditions are open and $\gamma = 1.4$ in both the tests. Initial conditions  
for 2D and 3D tests are given by Eq.\eqref{sb2d} and Eq.\eqref{sb3d}, respectively as,
 \begin{eqnarray}\label{sb2d}
  \left( \begin{array}{c} \rho, v_x,  v_y,  p\end{array}\right) = \left \{ \begin{array}{c} (1, 0, 0,  1) \hspace{0.4cm} 
  \text{if}  \hspace{0.2cm}  x > 0 \hspace{0.2cm}\text{and} \hspace{0.2cm}(x-0.3)^2 + y^2 \ge R^2 \hspace{0.7cm}\\ (1, 0, 0, 10)   \hspace{0.2cm} 
  \text{if} \hspace{0.2cm} x < 0 \hspace{5.5cm}\\ (0.1, 0, 0, 1)   \hspace{0.2cm} 
  \text{otherwise}  \hspace{5.1cm} \end{array}\right \}.  
  \end{eqnarray}

 \begin{eqnarray}\label{sb3d}
  \left( \begin{array}{c} \rho, v_x,  v_y, v_z,  p\end{array}\right) = \left \{ \begin{array}{c} (1, 0, 0, 0, 1) \hspace{0.4cm} 
  \text{if}  \hspace{0.2cm}  x > 0 \hspace{0.2cm}\text{and} \hspace{0.2cm}(x-0.3)^2 + y^2  + z^2 \ge R^2 \\ (1, 0, 0, 0, 10)   \hspace{0.2cm} 
  \text{if} \hspace{0.2cm} x < 0 \hspace{5.5cm}\\ (0.1, 0, 0, 0, 1)   \hspace{0.1cm} 
  \text{otherwise}  \hspace{5.1cm} \end{array}\right \}.  
  \end{eqnarray}
Here $R$ is chosen to be $0.2$ in both 2D and 3D tests. In figures~\ref{fig1-sc-2d} and~\ref{fig2-sc-2d} we 
show the snap-shots of color-coded contour plots of the 
density at time $t = 0.1$ and $t = 0.4$, respectively for the 2D shock-bubble interaction test.  
In figures~\ref{fig2-sc-2d} and~\ref{fig2-sc-3d} we 
show the snap-shots of color-coded contour plots of the 
density at time $t = 0.1$ and $t = 0.4$, respectively for the 3D shock-bubble interaction test in the plane $z = 0$. 
In both the tests, the 
results are found to be the same as reported in \cite{shock-cloud2D,shock-cloud3D}. 

We have thus shown the shock capturing nature and non-oscillatory behavior of the scheme in various nonlinear tests. In the next section we will 
talk about the advantage of the present fourth order implementation over third order \cite{kurganov-2000}. 
 
\section{Comparison with the third order central schemes}
In this section we provide a comparison between the present fourth order central scheme and 
the third order central scheme \cite{kurganov-2000} for the 2D Euler vortex problem, for the 3D 
blast wave problem and Kelvin-Helmholtz instability to give an estimate of numerical dissipation for smooth and non-smooth solutions.
We also make an estimate of the computational cost for the two schemes. 
It is to mention here that we use third order accurate averaged fluxes (as computed 
from the point-value fluxes in method B and method C of section III) to ensure the third-order accuracy \cite{kurganov-2000}. 

{\bf The 2D Euler vortex problem} -- We solve the 2D Euler vortex problem, described in the Section IV, for the present fourth order 
central scheme and the third order central scheme \cite{kurganov-2000}. 
Solutions are obtained at a resolution $96^2$  and density profiles, along the diagonal, are compared after $10$ periods $(t = 100)$ 
which are shown in Figure~\ref{fig10}.
It is found from this nonlinear test that the present fourth order scheme has a significantly smaller numerical dissipation as 
compared to the third order scheme \cite{kurganov-2000}. 

{\bf The 3D blast (explosion) test problem} -- The initial conditions for this test are same as described in the previous section. 
Here the solutions for both the schemes are obtained at time $t = 0.1$ at a resolution $80^3$. Figure~\ref{fig11} 
shows the 1D profile of the density $(\rho)$ along the intersection of planes 
$y = 0.5$ and $z = 0.5$ for both the schemes and reference solution is obtained at a high resolution $720^3$ using 
the fourth order scheme. 

{\bf Kelvin-Helmholtz (KH) instability} --  The initial conditions for this test are same as described in the previous section, however, 
wavelength of the perturbation $\lambda$ is chosen to be $1/3$ and evolution of the KH instability is 
shown from time $t = 1.5$ to $t = 6.0$ as color-coded contour plots of the density in all the figures. 
Figures~\ref{fig-kh-128-cweno3} and~\ref{fig-kh-256-cweno3} show the evolution of the KH instability obtained from the third order CWENO scheme 
at a resolution $128^2$ and $256^2$, respectively. Note here that the intrinsic behaviour of the KH instability are captured at a 
resolution $256^2$, however, at $128^2$ resolution these features have vanished due to higher numerical dissipation of the 
third order CWENO scheme. In contrast, as is shown in figure~\ref{fig-kh-128-cweno4} our fourth order CWENO scheme is 
able to capture these features even at $128^2$ resolution.      

{\bf Computational cost} --  In TABLE \ref{tab9} we compare the performance of the two schemes for 3D linear advection test 
and find that the third order scheme is on average $1.38$ times 
faster than present fourth order central scheme.   
\begin{center}
\begin{tabular}{ |p{1cm}|p{3cm}|p{3cm}| }
 \hline
NG & Third order & Fourth order \\
 \hline
$32^3$ & 1.0  & 1.36\\
$64^3$ & 1.0 &  1.41\\
$128^3$ &  1.0  & 1.41\\
$256^3$& 1.0  & 1.40\\
$512^3$& 1.0  & 1.32\\
 \hline
Mean & 1.0  & 1.38\\

 \hline
\end{tabular}
\captionof{table}{Performance test for the third and fourth order schemes where the run time is 
normalized with respect to the third order scheme.} \label{tab9}
\end{center}

Thus, we see that the implementation of the present fourth order central scheme 
involves an additional non-negligible but generally acceptable computational expense together with significantly reduced numerical dissipation 
when compared with the third order implementation \cite{kurganov-2000}.

\section{Conclusion}
A genuine multi-dimensional reconstruction for higher order schemes is complex and computationally 
more expensive than a dimension-by-dimension approach. Therefore, in this paper we have adopted 
a dimension-by-dimension CWENO approach to obtain a fourth order accurate central scheme. 
Different ways to employ a dimension-by-dimension CWENO approach have been discussed and  
the most efficient method has been applied to develop a fourth order CWENO scheme 
to solve multidimensional hyperbolic problems. 
The fourth order accuracy and the non-oscillatory property are confirmed in various multi-dimensional 
problems. 
The accuracy of the present dimension-by-dimension CWENO approach is found to be identical with 
truly multi-dimensional CWENO schemes.   
Moreover, the benefits of implementing the fourth order central scheme over third order central scheme have also been 
demonstrated by comparing the numerical dissipation and computational cost.

Fruitful discussions with J. Stone and P. Buchm\"uller are gratefully acknowledged and 
P.S.V. would also like to thank Tapan Chandra Adhyapak for useful suggestions.



\pagebreak

\begin{figure}
\includegraphics[height=9.5in,width=6.5in]{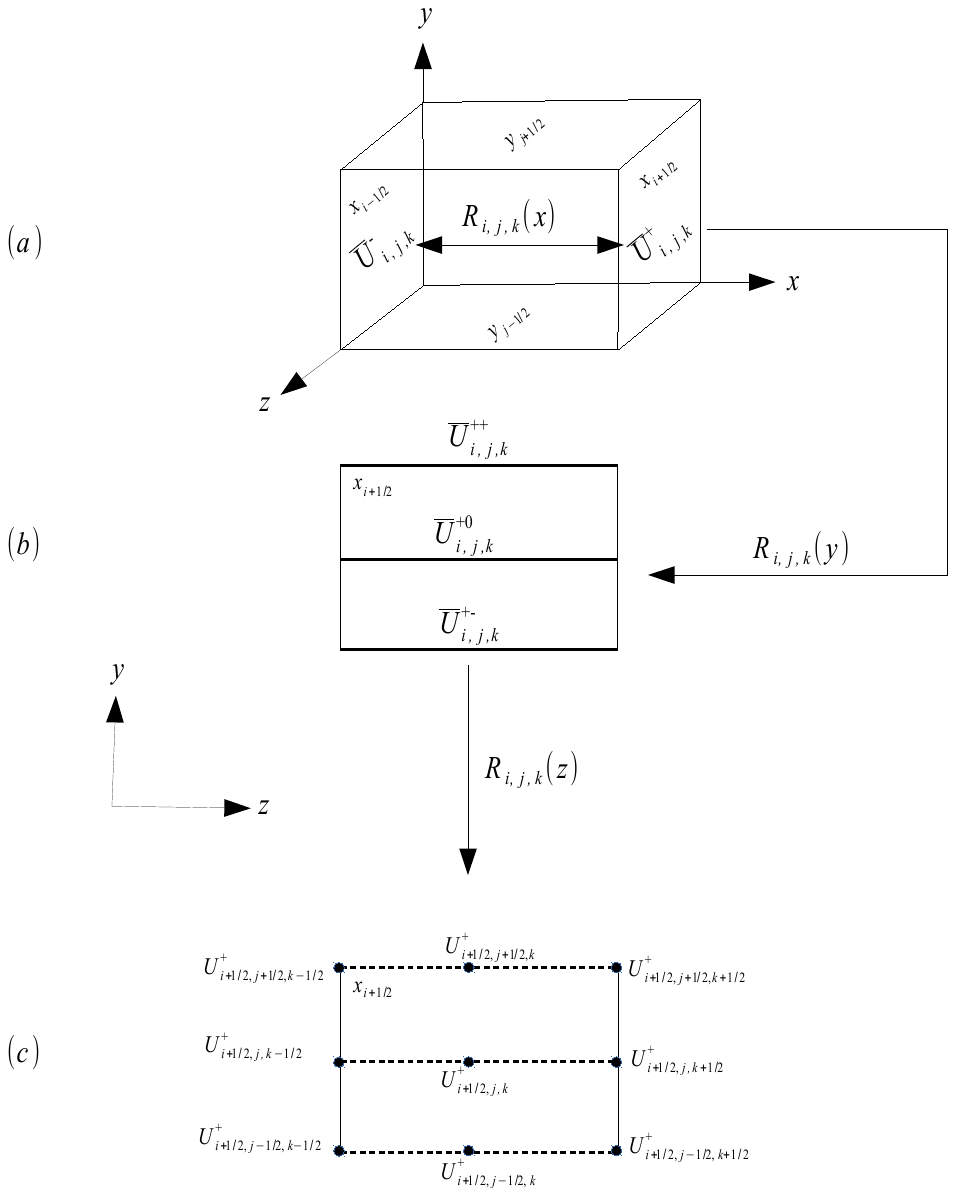}
{\vspace{-1.9in}\caption{ (a) Sketch of a 3D grid cell and reconstructed area-averages $\overline{\bf U}_{i,j,k}^{\pm}$ at the grid cell interfaces 
$x_{i\pm1/2}$ obtained from 
the 1D CWENO polynomial ${\bf R}_{i,j,k}(x)$ based on cell average $\overline{\bf U}_{i,j,k}$, (b) reconstructed three equi-distant edge-averages along 
the $y$-direction from the 1D CWENO polynomial ${\bf R}_{i,j,k}(y)$ based on area average $\overline{\bf U}_{i,j,k}^{+}$ at $x_{i+1/2}$, (c) 
reconstructed nine point values at $x_{i+1/2}$ 
from ${\bf R}_{i,j,k}(z)$ based on edge-averages in (b). 
}\label{fig1}}
\end{figure}
\vspace{-0.00in}

\clearpage

\begin{figure}
\includegraphics[height=5.5in,width=3.5in]{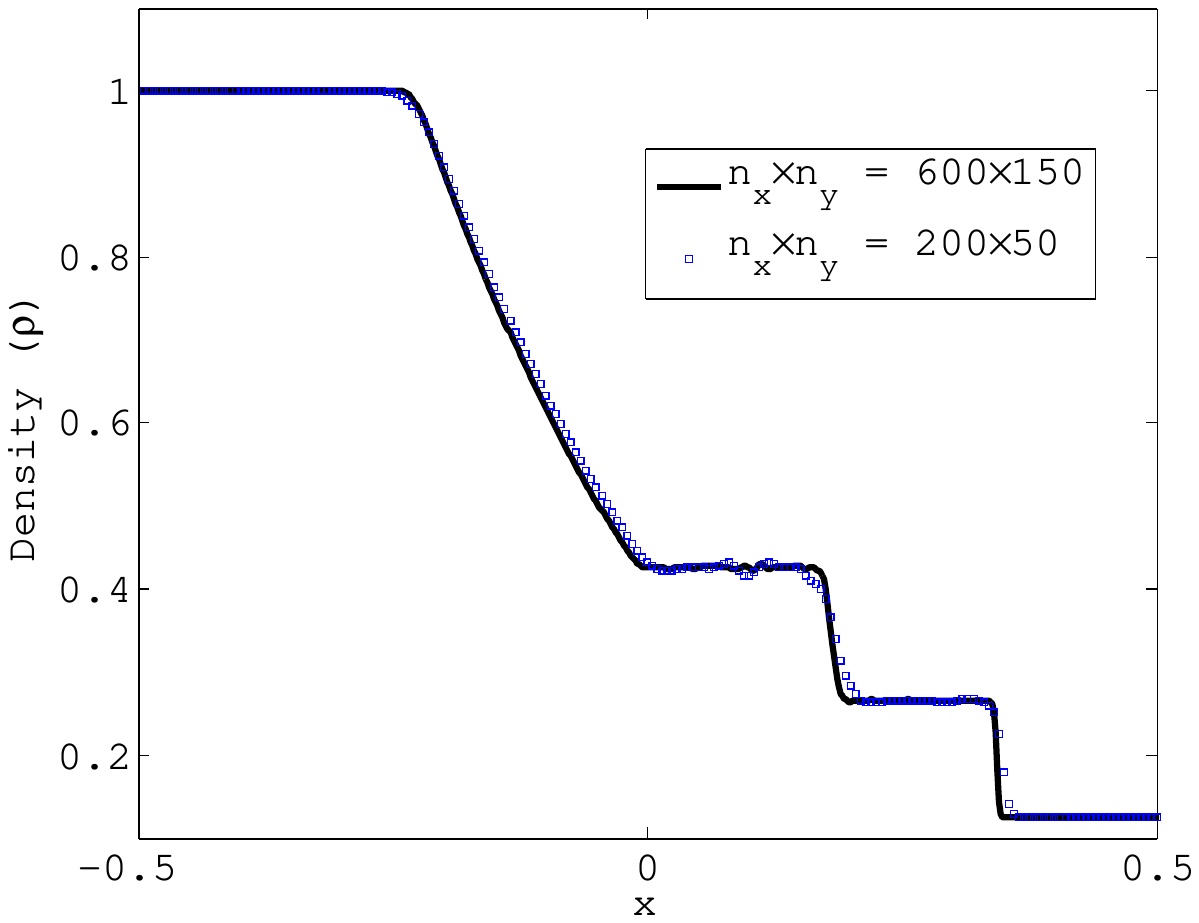}
{\vspace{-1.5in}\caption{Density $(\rho)$ along the $x$-direction for the oblique shock tube problem 
at two resolutions ($200\times50$) shown by 
`squares' and ($600\times150$) shown by `solid line' at time $t = 0.16632$. Here $n_x$ and $n_y$ are the number 
of grid points along the $x$ and $y$-directions, respectively.}\label{fig2}}  

\end{figure}

\begin{figure}
\includegraphics[height=4.8in,width=3.0in]{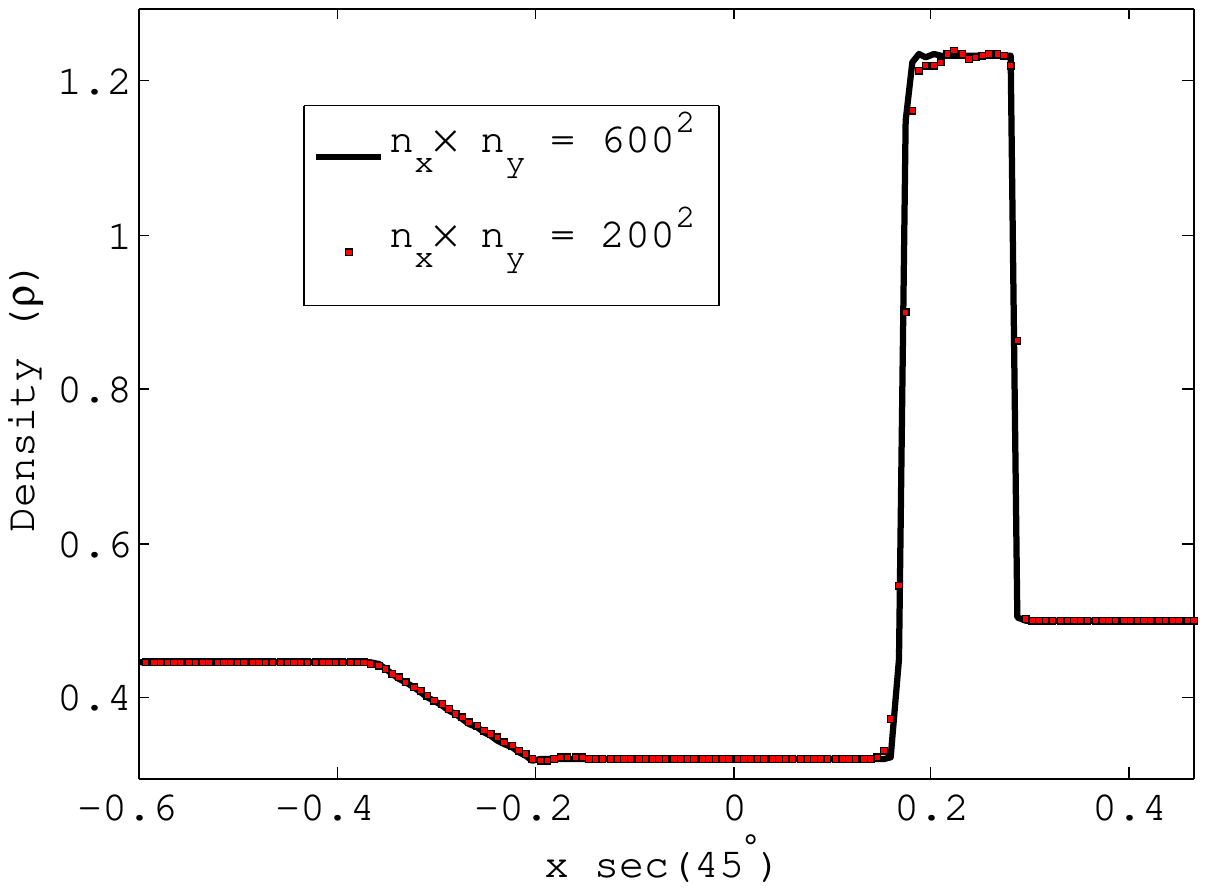}
{\vspace{-1.2in}\caption{ Solution of the oblique Lax problem : 1D profile of the density $(\rho)$ at an angle $45^{\circ}$ with 
the $x$-axis obtained at a resolution $200^2$ (red squares) and $600^2$ (solid line) at time $(t = 0.12)$.}\label{lax}}
\end{figure}

\begin{figure}
\includegraphics[height=5.2in,width=3.5in]{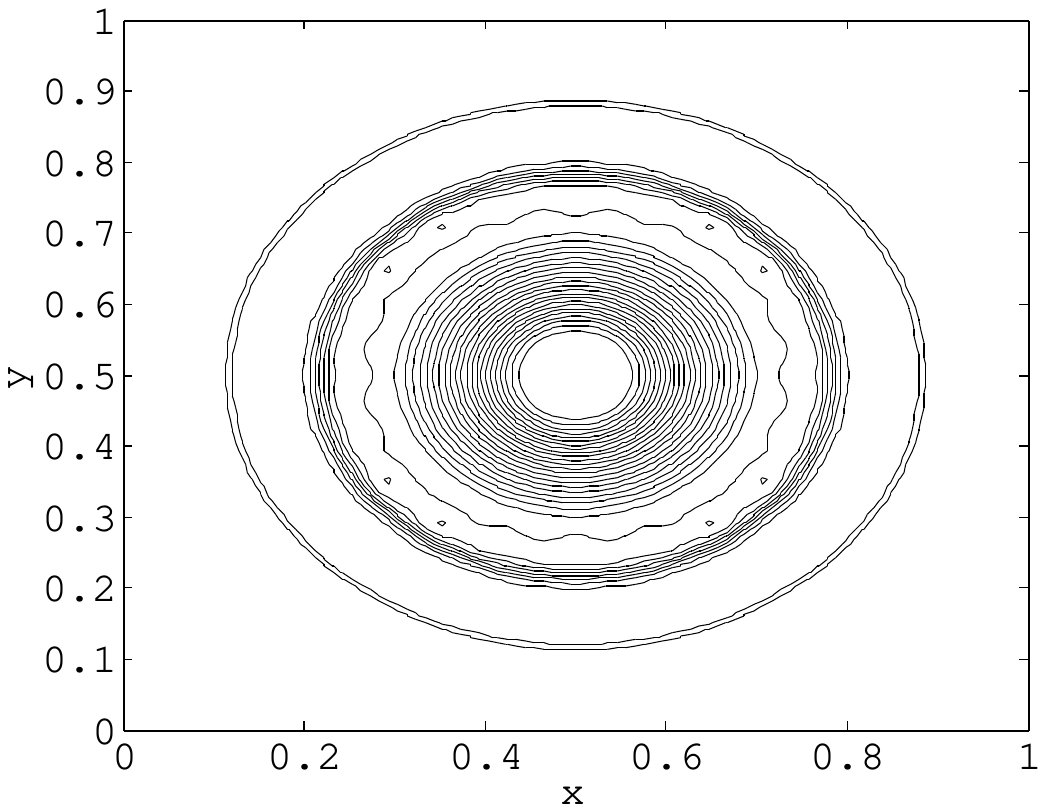}
{\vspace{-1.5in}\caption{Contour plot of the density ($\rho$) for the 2D blast wave problem at time $t = 0.1$ at a resolution $100^2$. }\label{fig3}}
\end{figure}

\begin{figure}
\includegraphics[height=4.5in,width=3.5in]{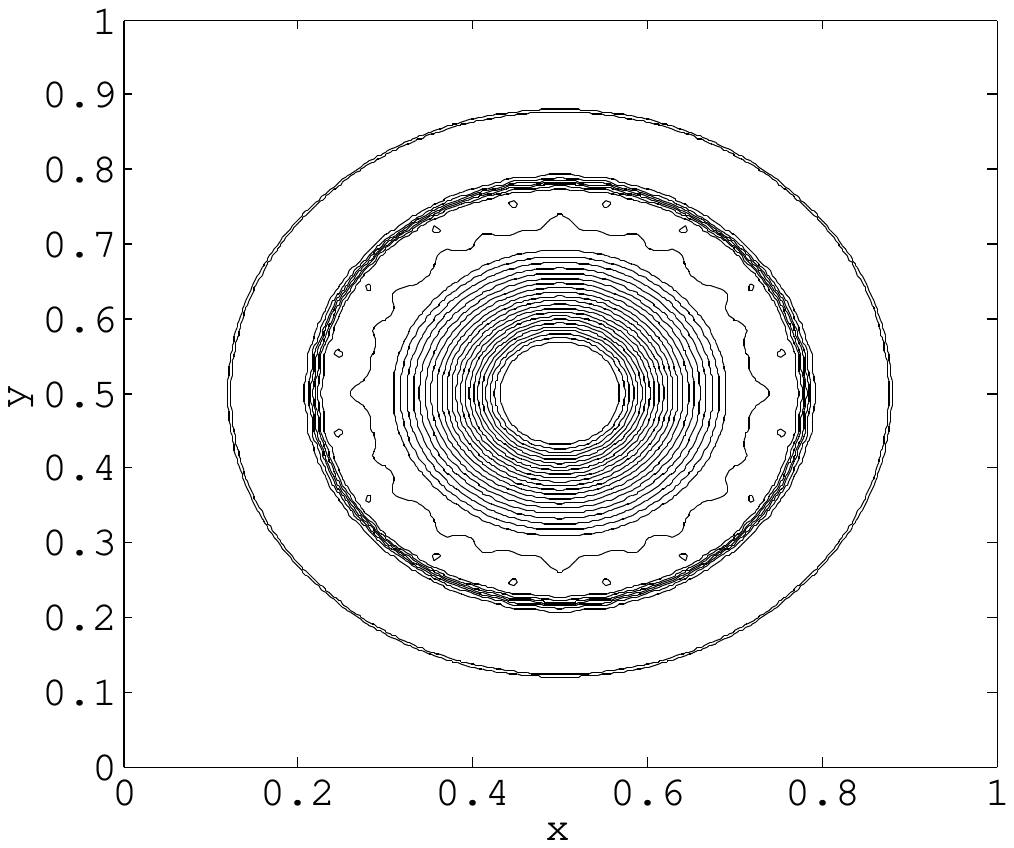}
{\vspace{-1.4in}\caption{Contour plot of the density ($\rho$) for the 2D blast wave problem at time $t = 0.1$ at a resolution $200^2$.  }\label{fig4}}
\end{figure}

\begin{figure}
\includegraphics[height=5.2in,width=3.5in]{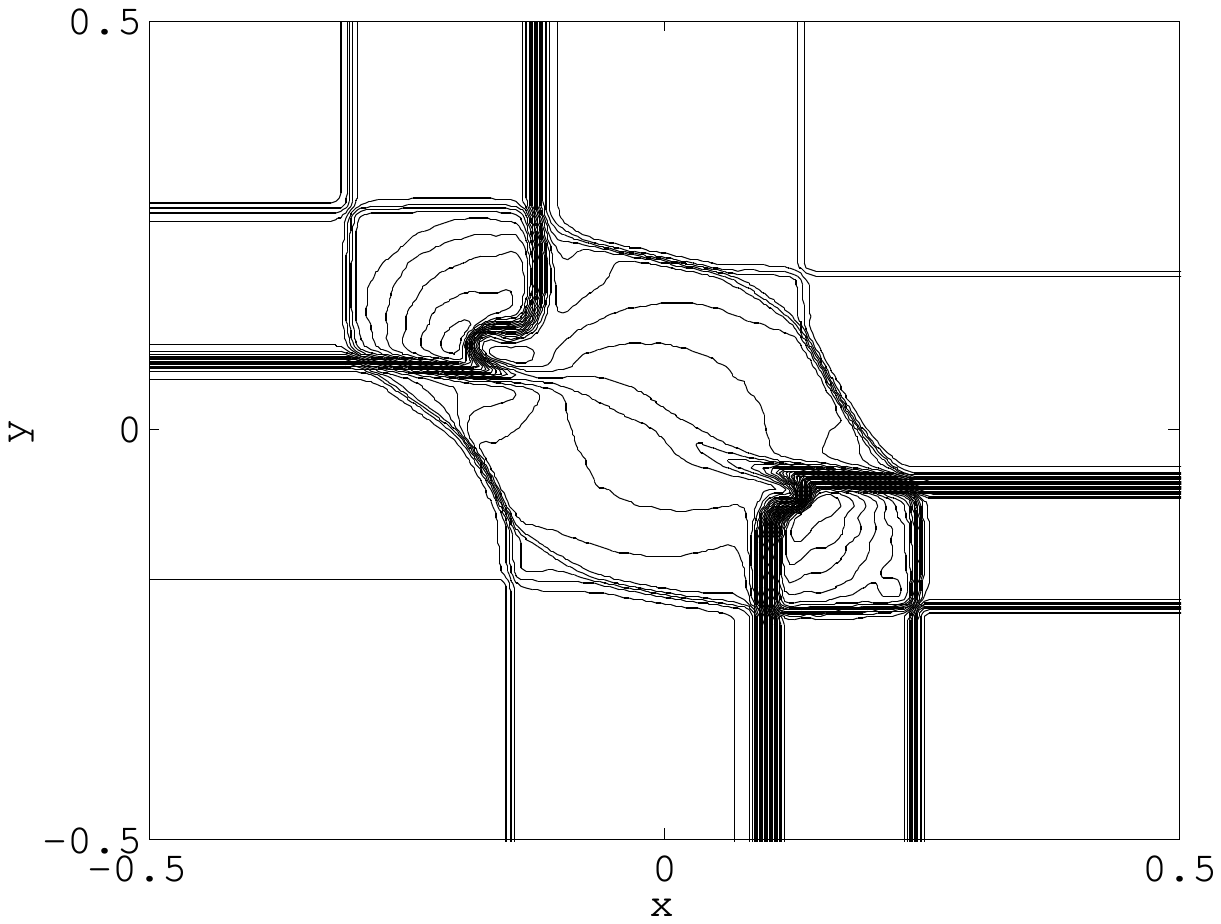}
{\vspace{-1.2in}\caption{ Contour plot of the density ($\rho$) for the 2D Riemann problem at time $t = 0.23$ at a resolution $200^2$. }\label{fig5}}
\end{figure}

\begin{figure}
\includegraphics[height=5.3in,width=3.5in]{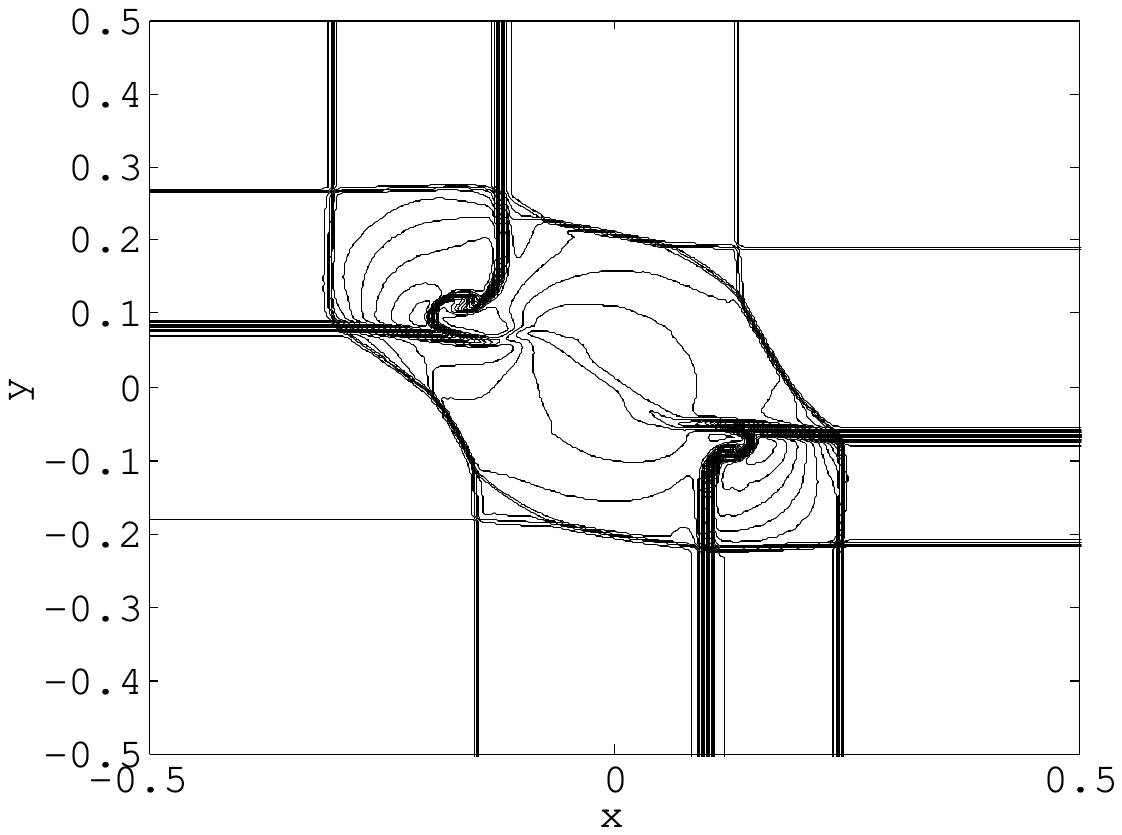}
{\vspace{-1.1in}\caption{  Contour plot of the density ($\rho$) for the 2D Riemann problem at time $t = 0.23$ at a resolution $400^2$. }\label{fig6}}
\end{figure}

\begin{figure}
\includegraphics[height=5.3in,width=3.5in]{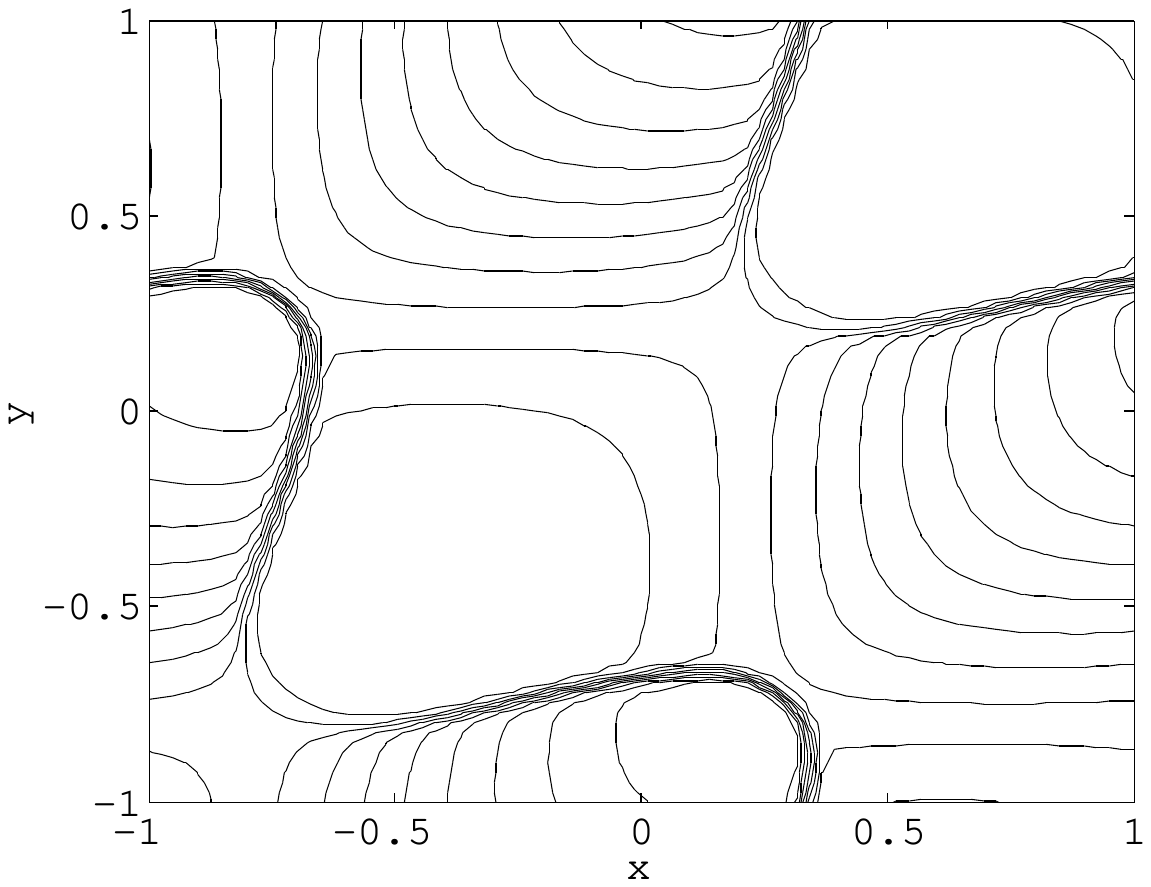}
{\vspace{-1.2in}\caption{Solution of the 3D Burgers' equation at time $t = 0.8$ : contour plot at the plane $z = 0$ at a resolution $80^3$. }\label{fig7}}
\end{figure}

\begin{figure}
\includegraphics[height=5.3in,width=3.5in]{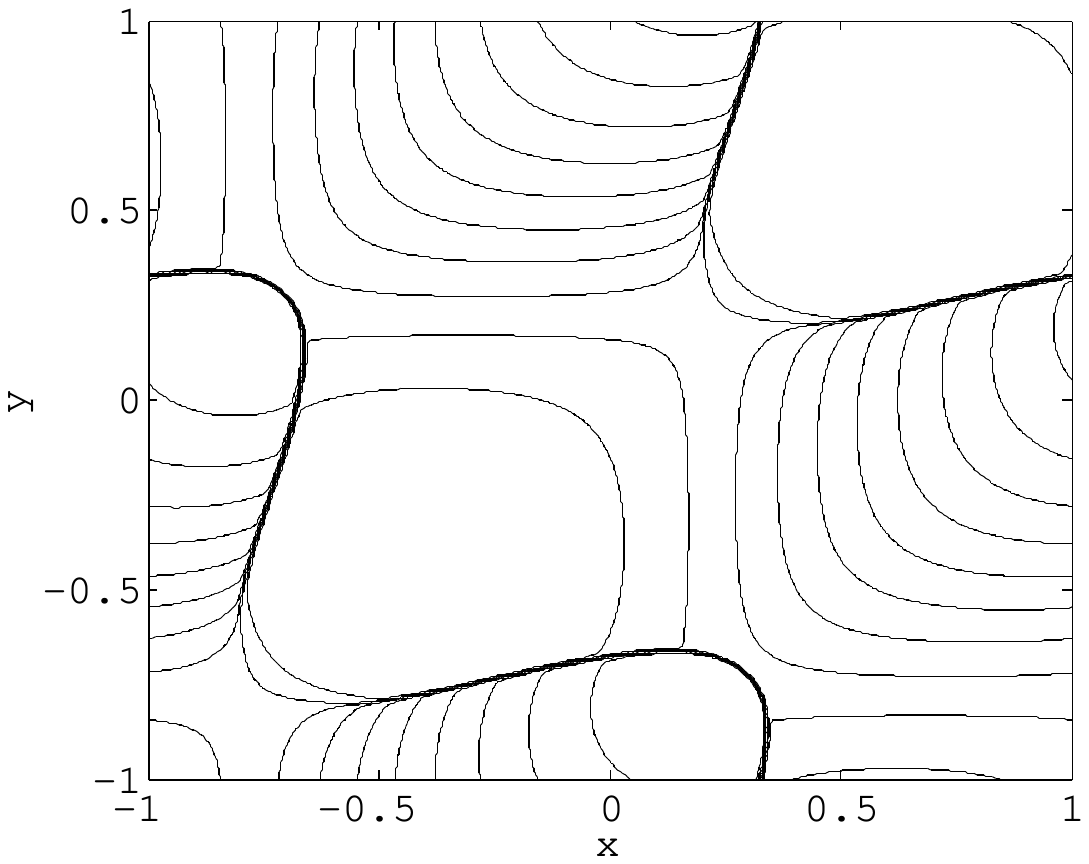}
{\vspace{-1.3in}\caption{ Solution of the 3D Burgers' equation at time $t = 0.8$ : contour plot at the plane $z = 0$ at a resolution $400^3$.}\label{fig8}}
\end{figure}

\begin{figure}
\includegraphics[height=4.5in,width=3.5in]{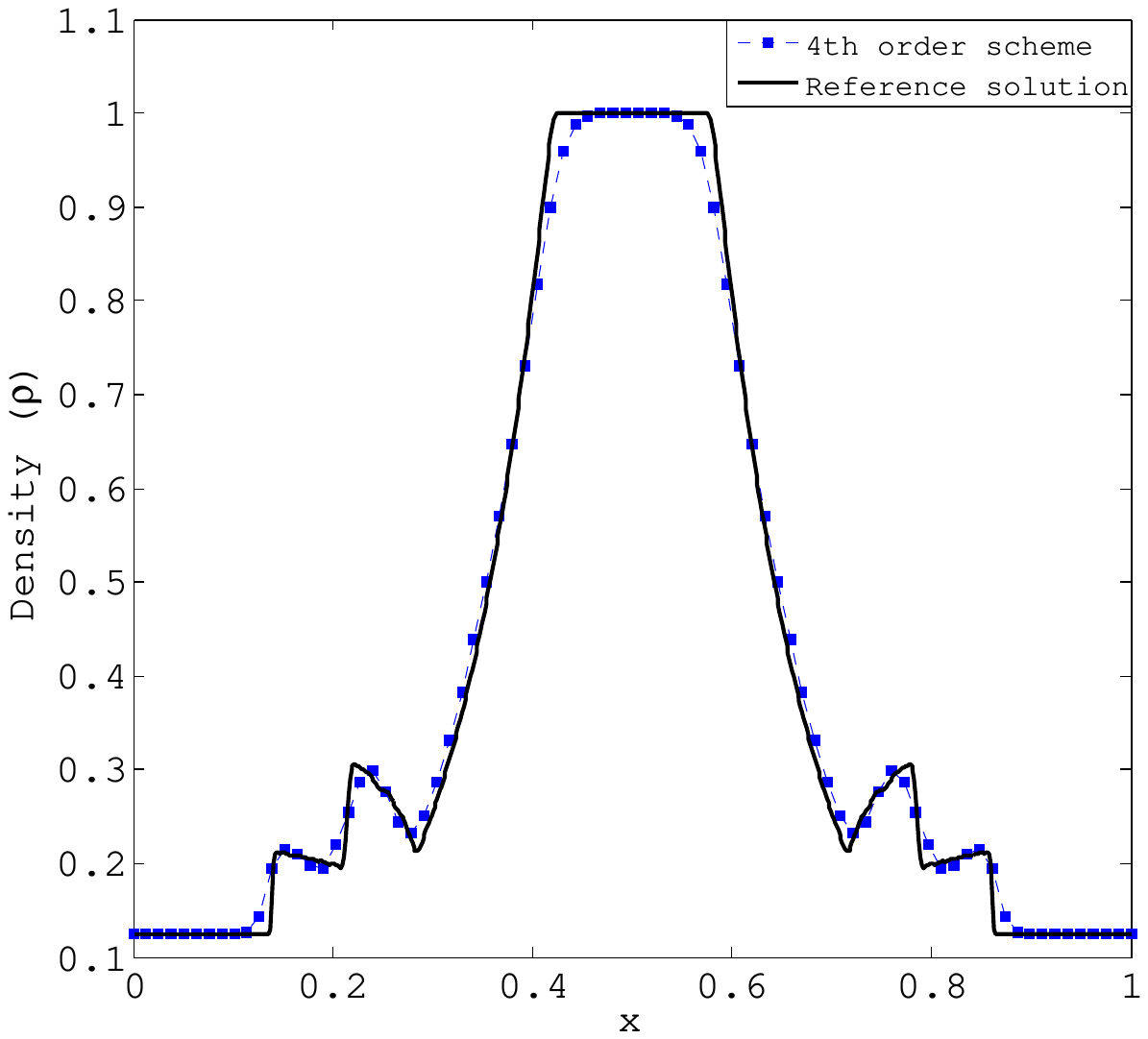}
{\vspace{-1.0in}\caption{Solution of the 3D blast wave problem at time $t = 0.1$ : 1D profile of the density $(\rho)$ along the intersection of planes 
$y = 0.5$ and $z = 0.5$ where the solid squares stand for the fourth order accurate solution at a resolution $80^3$ 
and solid line is the reference solution.}\label{fig9}}
\end{figure}

\begin{figure}
\includegraphics[height=5.5in,width=6.5in]{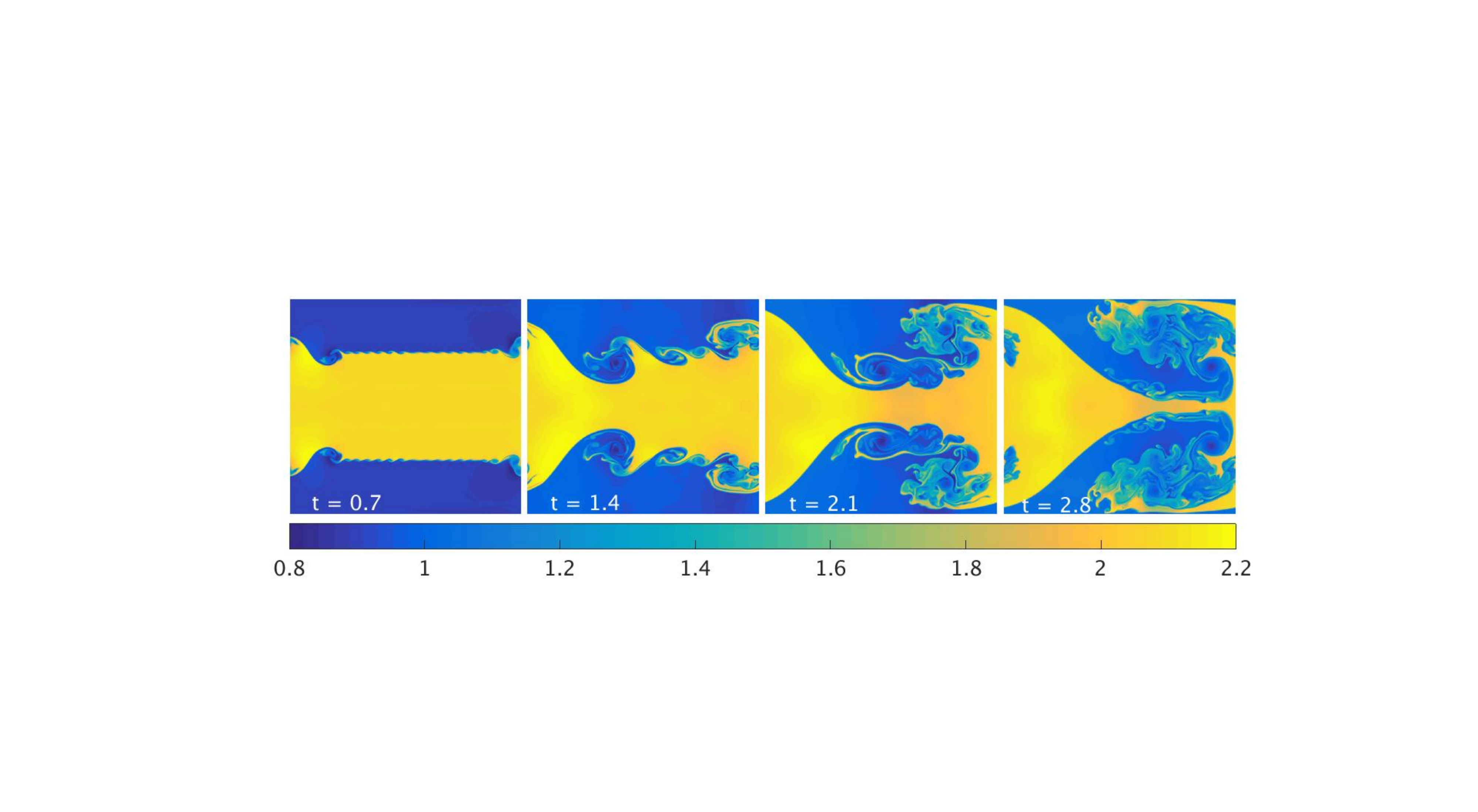} 
{\vspace{-1.8in}\caption{Snap-shot of Kelvin-Helmholtz instability at time $t = 0.7, 1.4, 2.1, 2.8$ : color-coded contour plots of the density $(\rho)$ 
 at a resolution $1024^2$.}\label{fig-kh-1024}}
\end{figure}

\begin{figure}
\includegraphics[height=4.5in,width=7.0in]{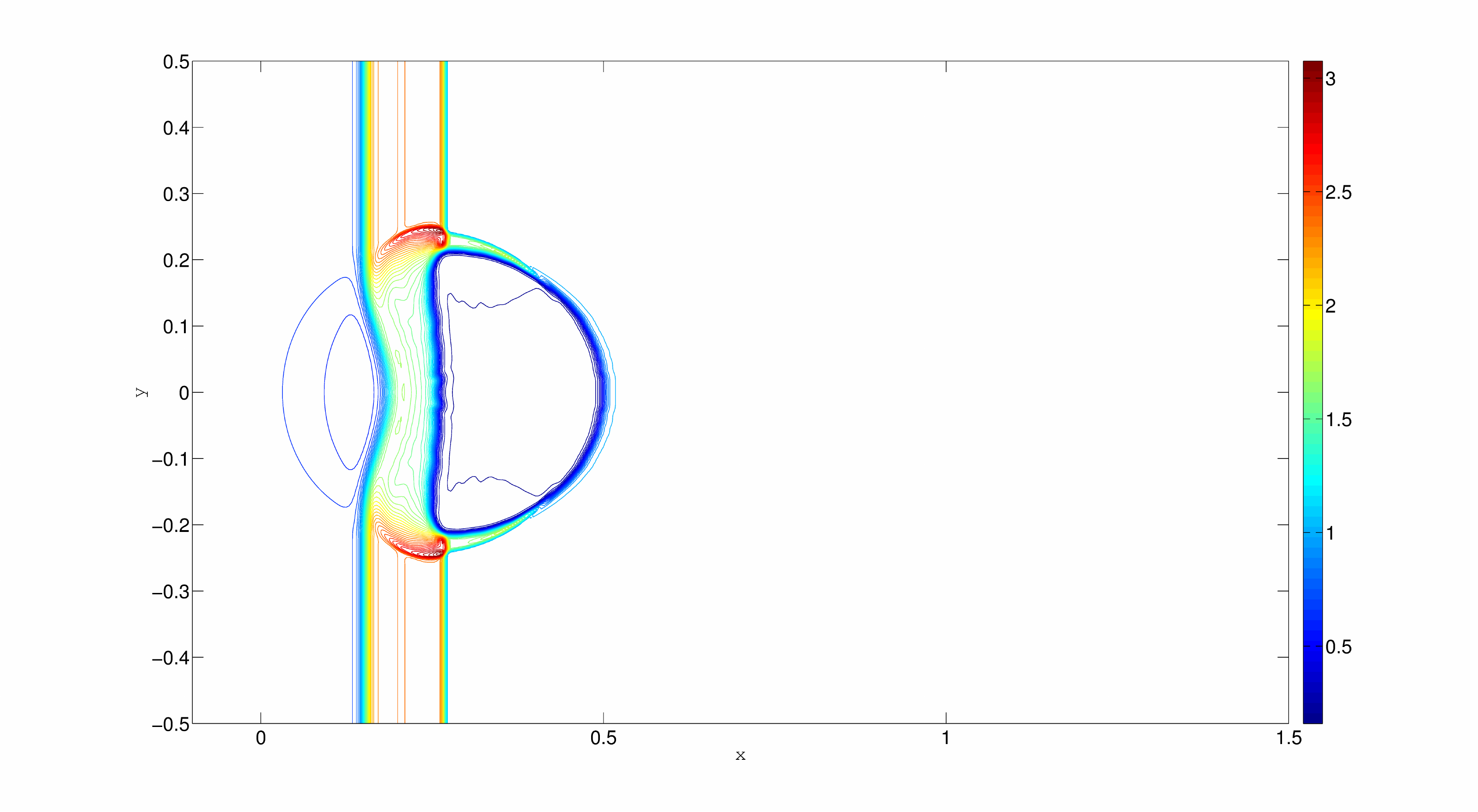} 
{\vspace{-0.4in}\caption{Snap-shot of a 2D shock-bubble interaction at time $t = 0.1$ : color-coded contour plots of the density $(\rho)$ 
 at a resolution $400^2$.}\label{fig1-sc-2d}}
\end{figure}

\begin{figure}
\begin{center} 
\includegraphics[height=4.5in,width=7.0in]{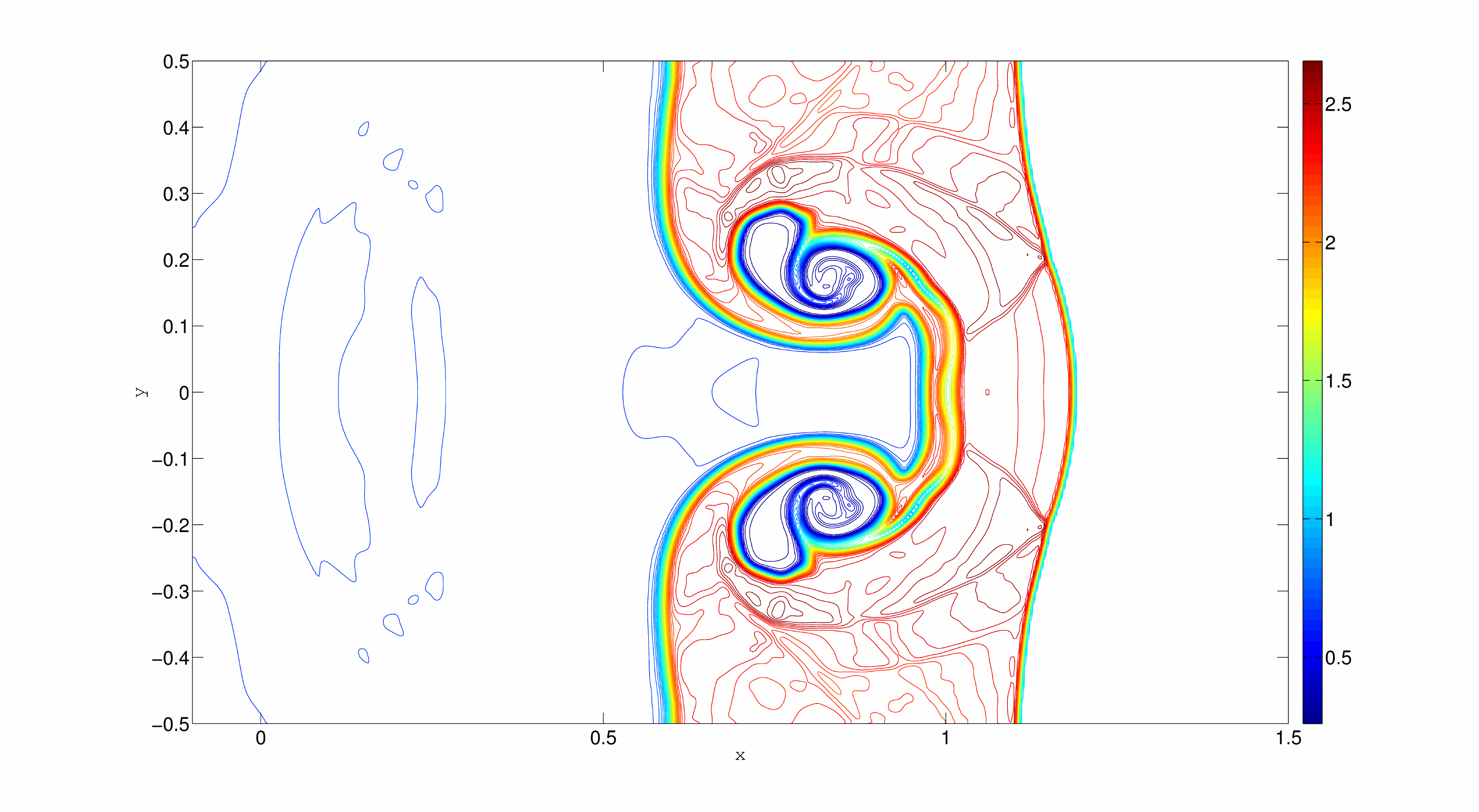} 
\end{center}
{\vspace{-0.4in}\caption{Snap-shot of a 2D shock-bubble interaction at time $t = 0.4$ : color-coded contour plots of the density $(\rho)$ 
 at a resolution $400^2$.}\label{fig2-sc-2d}}
\end{figure}

\begin{figure}
\includegraphics[height=4.5in,width=7.0in]{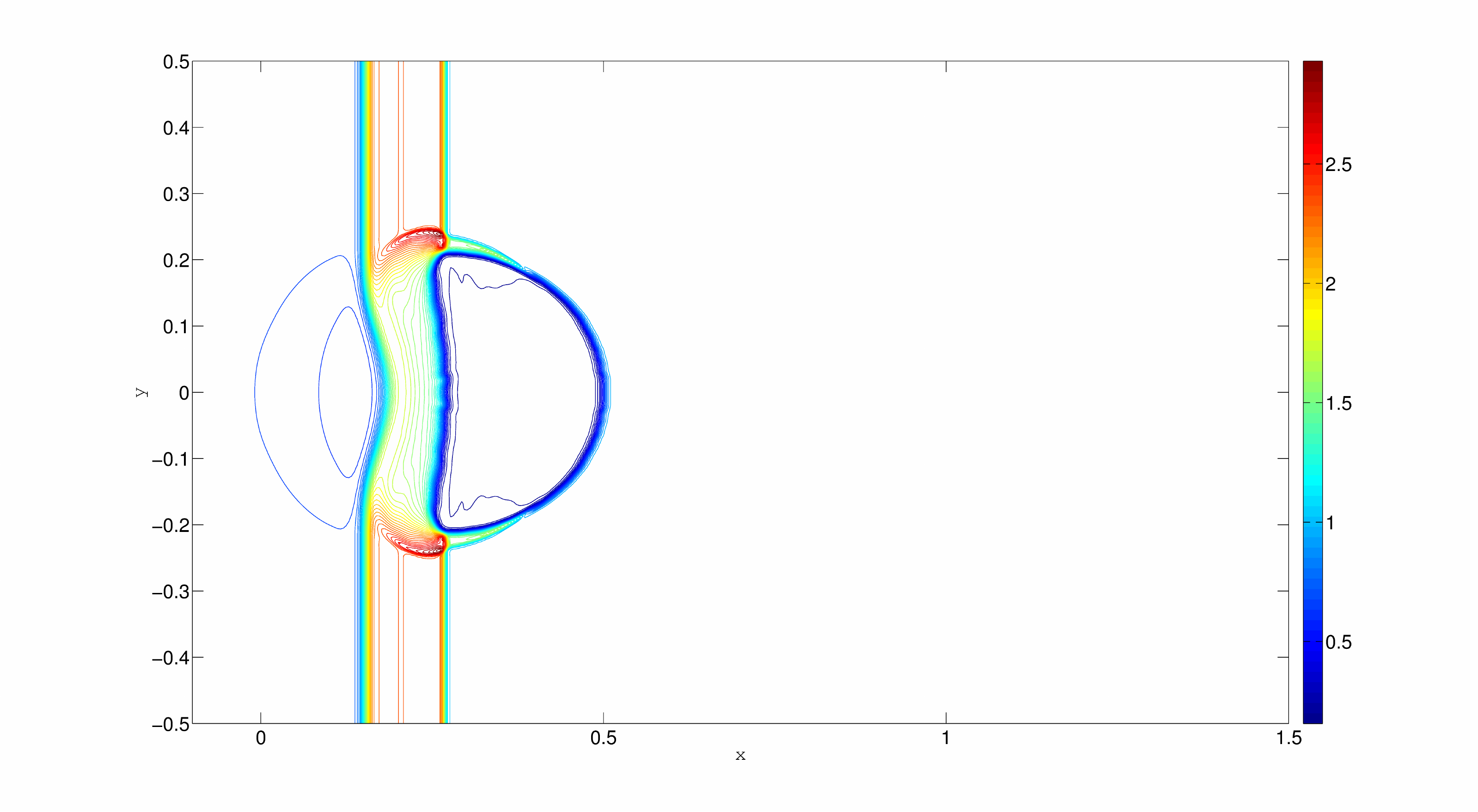} 
{\vspace{-0.4in}\caption{Snap-shot of a 3D shock-bubble interaction at time $t = 0.1$ : color-coded contour plots of the density $(\rho)$ 
 at a resolution $400^3$ in the plane $z=0$.}\label{fig1-sc-3d}}
\end{figure}

\begin{figure}
\includegraphics[height=4.5in,width=7.0in]{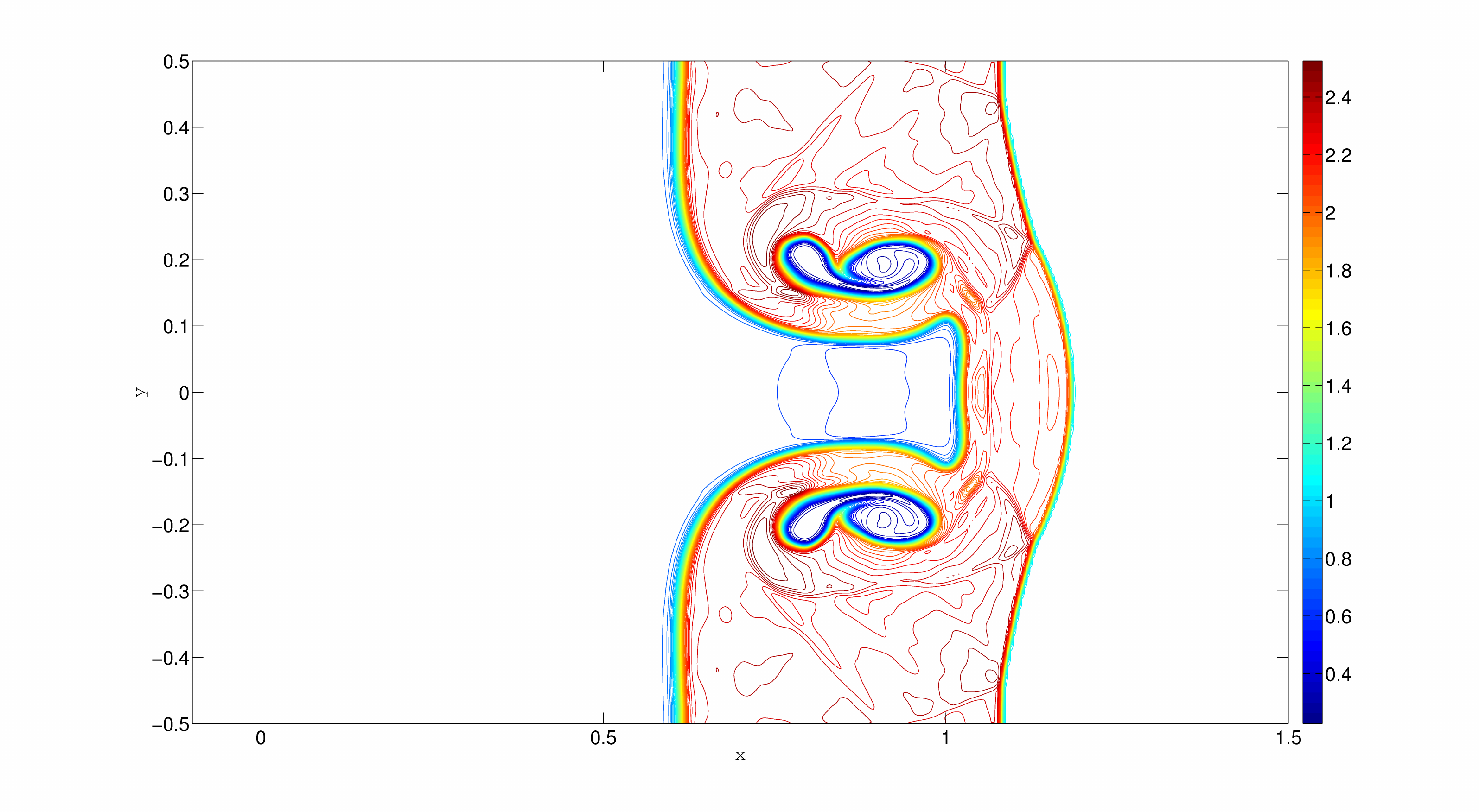} 
{\vspace{-0.4in}\caption{Snap-shot of a 3D shock-bubble interaction at time $t = 0.4$ : color-coded contour plots of the density $(\rho)$ 
 at a resolution $400^3$ in the plane $z=0$.}\label{fig2-sc-3d}}
\end{figure}

\begin{figure}
\includegraphics[height=6.0in,width=4.5in]{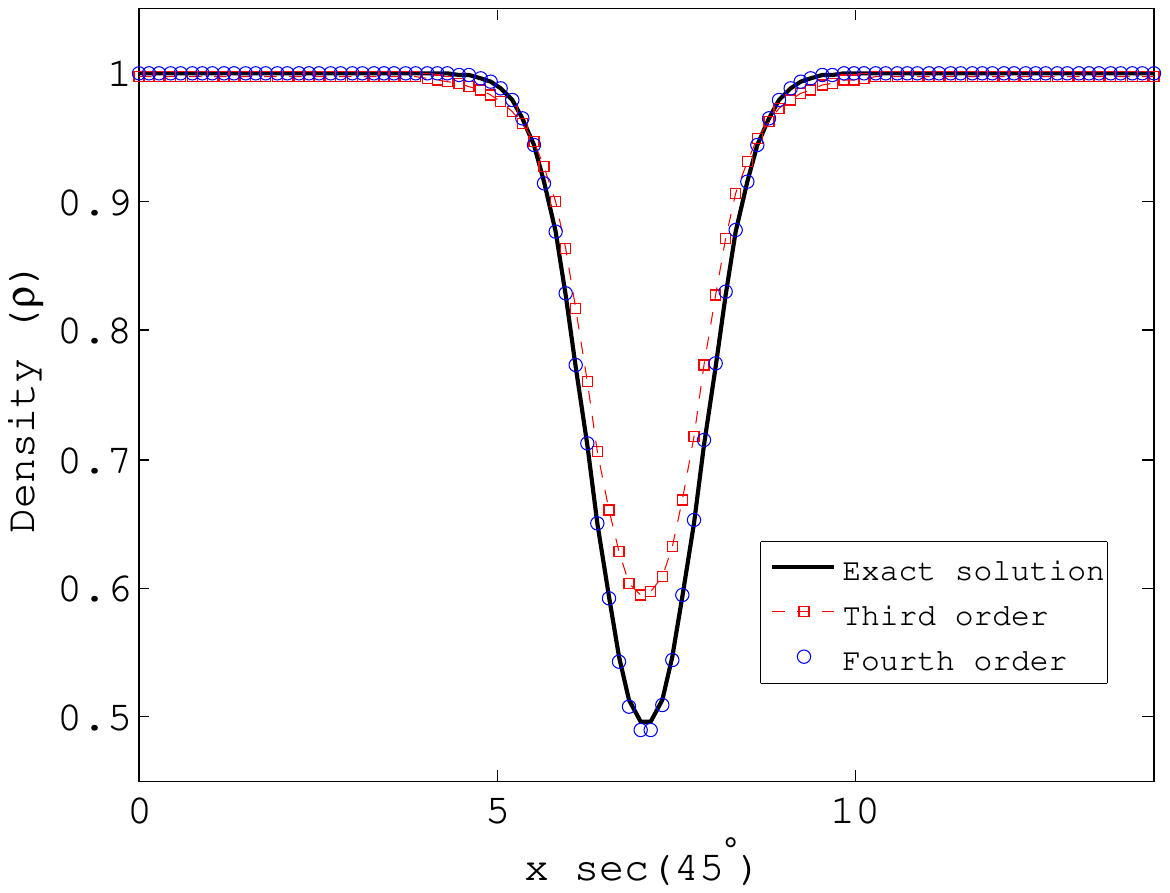}
{\vspace{-1.8in}\caption{ Solution of the 2D vortex problem : 1D profile of the density $(\rho)$ at an angle $45^{\circ}$ with 
the $x$-axis obtained at a resolution $96^2$ after $10$ periods $(t = 100)$ where squares represent the third order central scheme, circles stand 
for the present fourth order central scheme and the solid line `-' is the exact solution. }\label{fig10}}
\end{figure}

\begin{figure}
\includegraphics[height=4.5in,width=3.5in]{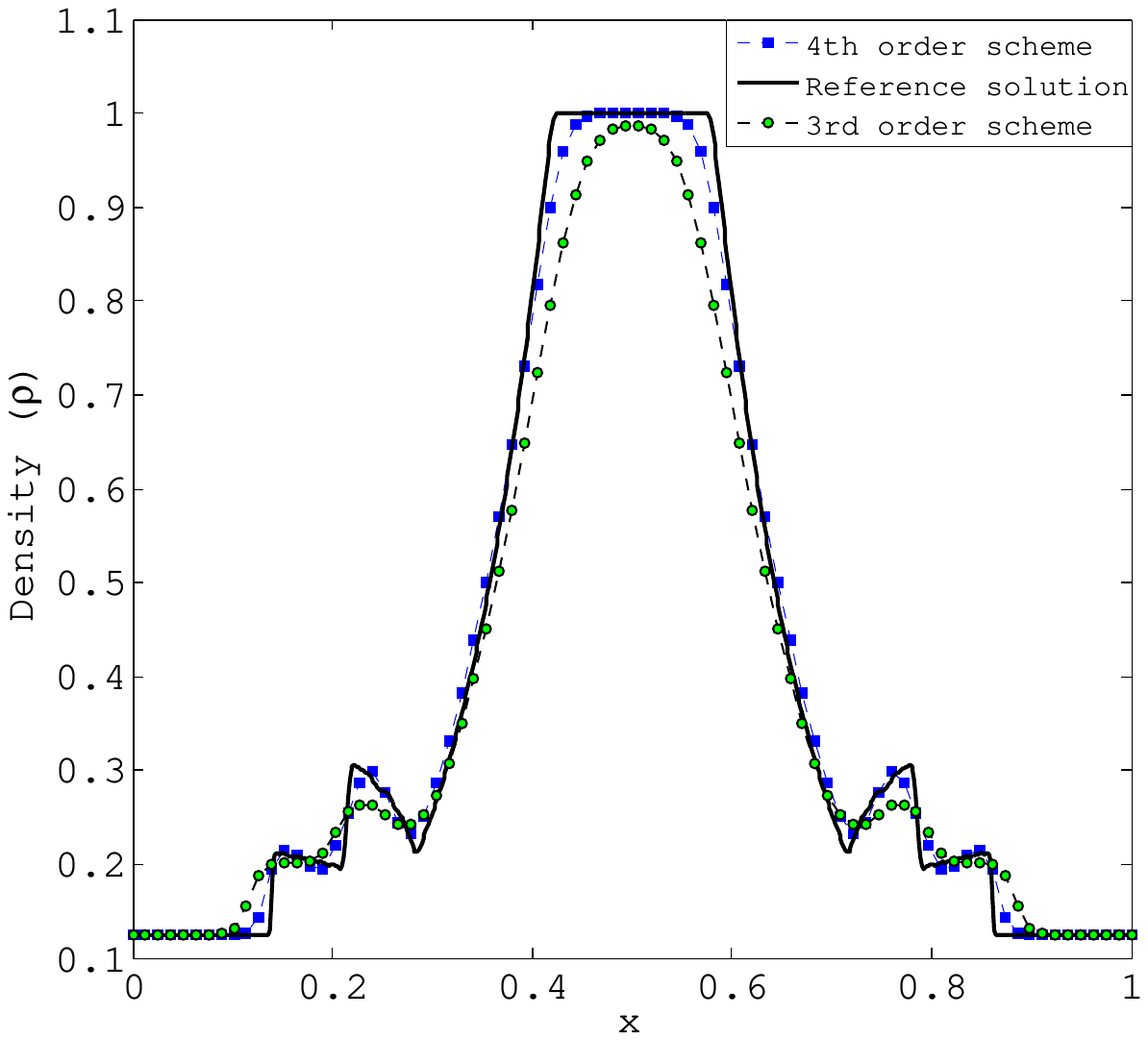}
{\vspace{-1.0in}\caption{Solution of the 3D blast wave problem at time $t = 0.1$ at a resolution $80^3$: 
1D profile of the density $(\rho)$ along the intersection of planes 
$y = 0.5$ and $z = 0.5$ where the solid squares stand for the fourth order accurate solution, circles depict the 
third order accurate solution and the solid line `-' is the reference solution.}\label{fig11}}
\end{figure}

\begin{figure}
\includegraphics[height=5.5in,width=6.5in]{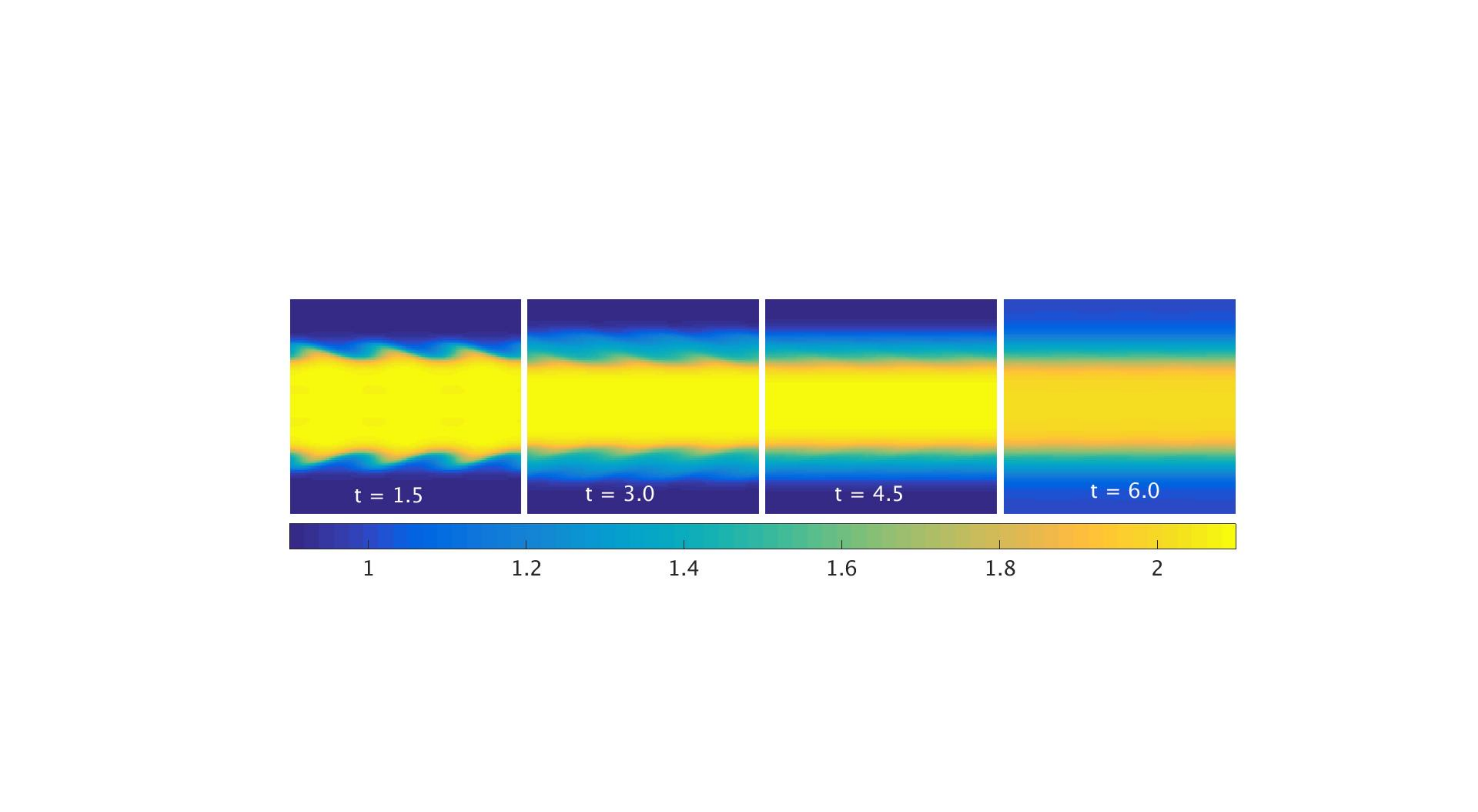} 
{\vspace{-1.8in}\caption{Evolution of Kelvin-Helmholtz instability at time $t = 1.5, 3.0, 4.5, 6.0$: color-coded contour plots of the density $(\rho)$ 
 at a resolution $128^2$ using third order CWENO scheme.}\label{fig-kh-128-cweno3}}
\end{figure}

\begin{figure}
\includegraphics[height=5.5in,width=6.5in]{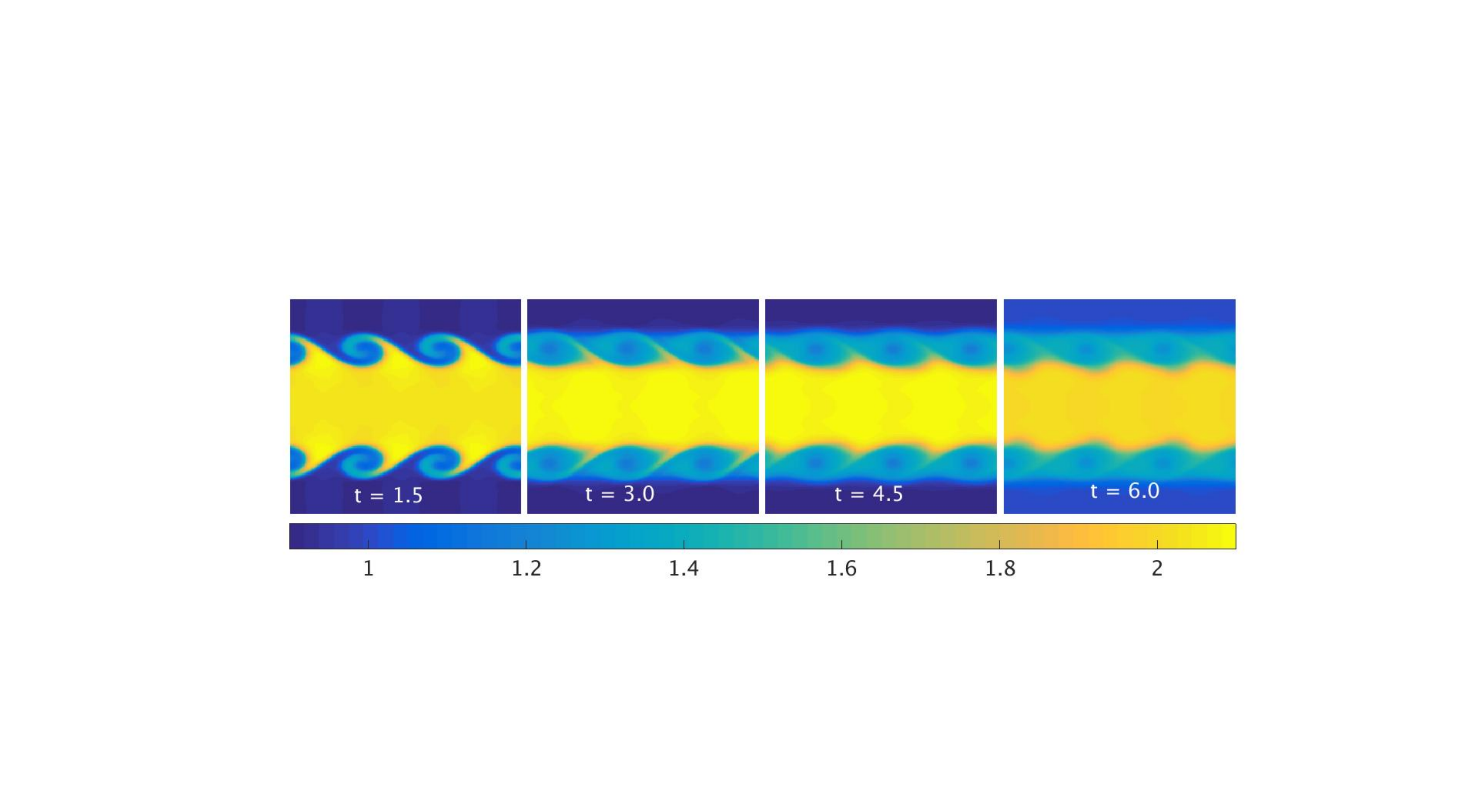} 
{\vspace{-1.8in}\caption{Evolution of Kelvin-Helmholtz instability at time $t = 1.5, 3.0, 4.5, 6.0$ : color-coded contour plots of the density $(\rho)$ 
 at a resolution $256^2$ using third order CWENO scheme.}\label{fig-kh-256-cweno3}}
\end{figure}

\clearpage

\begin{figure}
\includegraphics[height=5.5in,width=6.5in]{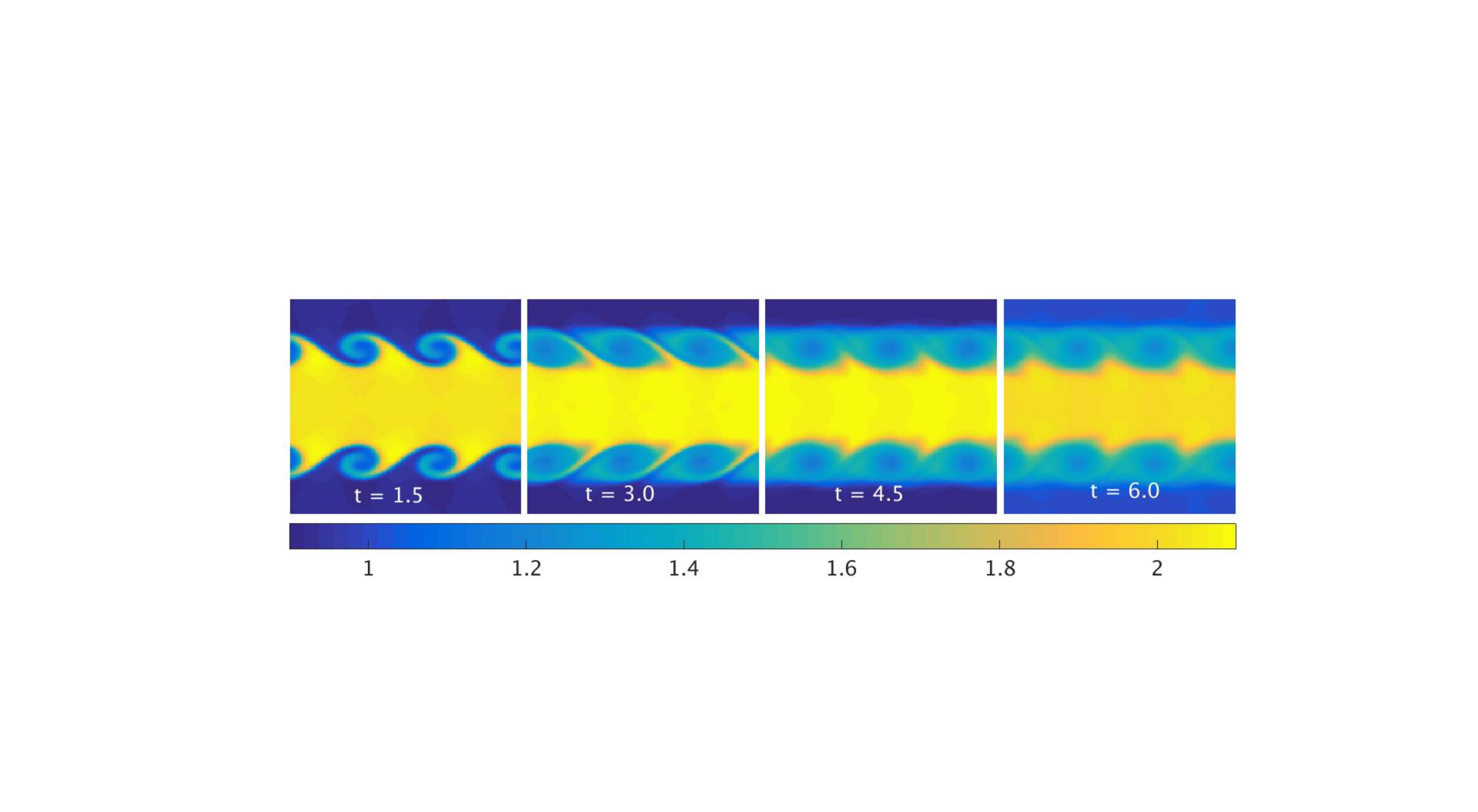} 
{\vspace{-1.8in}\caption{Evolution of Kelvin-Helmholtz instability at time $t = 1.5, 3.0, 4.5, 6.0$: color-coded contour plots of the density $(\rho)$ 
 at a resolution $128^2$ using fourth order CWENO scheme.}\label{fig-kh-128-cweno4}}
\end{figure}

\end{document}